\documentclass{aa}

\usepackage{amsmath}
\usepackage{physics}
\usepackage{graphicx}
\usepackage[labelfont=bf]{caption}
\usepackage{subfig}
\usepackage{rotating}
\usepackage{txfonts}
\usepackage{makecell}
\usepackage{multirow}
\usepackage{hyperref}
\usepackage[dvipsnames]{xcolor}
\hypersetup{
    colorlinks,
    linkcolor={blue!80!black},
    citecolor={blue!80!black},
    urlcolor={blue!80!black}
}
\usepackage{natbib}
\bibpunct{(}{)}{;}{a}{}{,}
\usepackage{nicefrac}
\usepackage{dblfloatfix}
\usepackage{boldline}

\newcommand{\Msun}{\textrm{M}$_\mathrm{\odot}$}
\newcommand{\Mearth}{\textrm{M}$_\mathrm{\oplus}$}
\newcommand{\Mj}{\textrm{M}$_\mathrm{J}$}

\newcommand{\abin}{$a_{\text{bin}}$}
\newcommand{\ebin}{$e_{\text{bin}}$}

\begin{document} 

 \title{The PAIRS project: a global formation model for planets in binaries}
   \subtitle{II. Gravitational perturbation effects from secondary stars}

   \author{Arianna Nigioni  
          \inst{\ref{unige}}
          \and
          Julia Venturini \inst{\ref{unige}}
          \and
          Emeline Bolmont \inst{\ref{unige}}
          \and
          Diego Turrini \inst{\ref{oato},\ref{inaf}}
          \and
          Yann Alibert \inst{\ref{unibe1},\ref{unibe2}}
          \and
          Alexandre Emsenhuber \inst{\ref{unibe1}}
          }

   \institute{Observatoire astronomique de l’Université de Genève, Chemin Pegasi 51, CH-1290 Versoix, Switzerland \label{unige}
   \and
    INAF - Osservatorio Astrofisico di Torino, via Osservatorio 20, 10025, Pino Torinese, Italy \label{oato}
    \and
    INAF-IAPS, Via Fosso del Cavaliere 100, 00133, Rome, Italy\label{inaf}
    \and
    Division of Space Research and Planetary Sciences, Physics Institute, University of Bern, Gesellschaftsstrasse 6, 3012 Bern, Switzerland \label{unibe1}
    \and 
    Center for Space and Habitability, University of Bern, Gesellschaftsstrasse 6, 3012 Bern, Switzerland \label{unibe2}
     }

   \date{}
    \titlerunning{The PAIRS project II: a global formation model for planets in binaries}
    \authorrunning{Arianna Nigioni et al.}
 
\abstract
{Roughly half of Sun-like stars have at least one stellar companion, whereas it is widely assumed that most known exoplanets orbit single stars, largely due to observational biases. However, astrometric surveys, direct imaging, and speckle interferometry are steadily increasing the number of confirmed exoplanets in binaries. A stellar companion introduces additional effects, such as circumstellar disk truncation and gravitational perturbations, which can strongly impact planet formation. While global planet formation models (e.g., Bern model) have been broadly applied to single stars, modeling S-type binaries requires key modifications to capture these effects.}
{This study extends the Bern model by incorporating the gravitational influence of a stellar companion into its N-body integrator, allowing us to quantify how this perturbation affects planetary formation and final system architecture across a range of binary configurations. By comparing binary and single-star systems under identical initial conditions, we can assess the specific impact of binary-induced dynamics.}
{We modified the Bern model’s N-body integrator to include secondary star perturbations and ran three sets of simulations: (i) a grid of in situ single-embryo cases to quantify gravitational effects; (ii) formation simulations with and without migration to compare outcomes with single-star analogs; and (iii) multi-embryo runs to evaluate impacts on multi-planetary systems.}
{Planets forming beyond half the host star’s Hill radius are much more likely to become unbound (i.e., in about six out of seven cases), especially in systems with high binary eccentricity. Even within stable zones, growth is suppressed by both reduced material availability (due to disk truncation) and increased eccentricity from stellar perturbations. Multi-embryo simulations have shown that these perturbations tend to reshape architectures even in dynamically stable regions.}
{Both disk truncation and stellar perturbations must be included to model planet formation in S-type binaries accurately. Neglecting either one will end up misrepresenting planetary growth and survival. Finally, population synthesis studies will be key in making statistical comparisons with observed systems.}


   \maketitle

\section{Introduction}

Around half of Sun-like stars are found in multiple star systems \citep[e.g.,][]{DuquennoyMayor1991,Raghavan2010}, yet the majority of the almost 6000 confirmed exoplanets reported in the \href{https://exoplanetarchive.ipac.caltech.edu/}{NASA Exoplanet Archive} have been found around single main sequence stars, mainly via transit and radial velocity methods. Overall, about 20\% have been reported to be in a binary star system \citep[e.g.,][]{ThebaultBonanni2025, Fontanive2021, Gonzales2024, MichelMugrauer2024}. The low proportion of exoplanets observed in binaries is biased by the fact that past surveys avoided observing binaries given the technical difficulties when targeting these systems. In recent years, however, astrometric surveys such as Gaia, along with ground-based follow-ups with direct imaging and speckle interferometry, have demonstrated the ability to detect bound companions to known planet-hosting stars, leading to an increase in the number of known exoplanets in binaries (e.g., \citealt{Mugrauer2020,Mugrauer2021,Mugrauer2022,Mugrauer2023,Mugrauer2025,Behmard2022}).

Planets in binaries can be found in two possible architectures: the S-type, where the planets orbit only one of the binary components, and the P-type, where the planets orbit around the centre of mass of the two stars \citep{Dvorak1986}. In this work, we focus on S-type systems. 
Modeling planet formation in these binaries and thereby enabling a statistical comparison with the observed S-type population requires a global formation model. One of the most detailed global model is the Bern model of planet formation and evolution \citep{Alibert2004,Alibert2005,Mordasini2012a,Mordasini2012b, Alibert2013,Fortier2013, NGPPS1}, which was developed for more than two decades to model planet formation and evolution around single stars. During the formation phase, it self-consistently computes the structure and evolution of the gas disk, the accretion of gas and solids by the planetary embryos, the interactions among planets via N-body integration, and the gas-planet interaction which leads to gas-driven migration. After the formation phase, it follows the long-term evolution of each planet by evaluating change in planetary masses due to atmospheric escape and change of their orbits due to stellar tides. The Bern model can follow the formation of planets through planetesimal accretion, while accounting for the dynamics of the planetesimal disk \citep{NGPPS1}, but can also simulate planetary growth through pebble accretion \citep{Brugger2018,Brugger2020} in the context of a two-population model that accounts for dust evolution and planetesimal formation \citep{Voelkel2020}.  

In the presence of a bound stellar companion, however, we have to account for additional physical effects. When studying planet formation around one star in a binary system, the circumstellar disk is truncated due to tidal forces coming from the companion star. This truncation limits the material available to form planets and, particularly in close binaries, introduces significant differences compared to single-star systems \citep[e.g.,][]{Papaloizou1977,Eggleton1983,Artymowicz1994,Pichardo2005,KleyNelson2007,Alexander2011,Manara2019,Zagaria2021,Ronco2021}. Additionally, we have to consider the gravitational influence of the secondary star throughout the entire planet formation process via N-body integration \citep{Chambers02}. 

A wide range of studies has investigated planet formation in S-type binaries, particularly focusing on systems such as $\alpha$ Centauri and $\gamma$ Cephei. Some works have modeled the late stages of planet formation, focusing on the accumulation of lunar-sized embryos into terrestrial planets via N-body simulations \citep[e.g.,][]{Barbieri2002, Quintana2002, Quintana2003, Lissauer2004, Thebault2004, Turrini2005, Turrini2006, Quintana2006, HaghighipourRaymond2007, Quintana2008}. There have also been dedicated efforts to explore how planetesimal growth and accretion are modified in the presence of a stellar companion, both in general cases and in specific binary systems. Starting from the early works of \citet{Heppenheimer1978} and \citet{Whitmire1998}, who used three-body calculations to show how stellar perturbations can increase the relative velocities of planetesimals, potentially leading to fragmentation, progressively more complex models have been developed \citep[e.g.,][]{MarzariScholl2000, Thebault2004, Thebault2006, Paardekooper2008, XieZhou2008, Thebault2008, Thebault2009, PaardekooperLeinhardt2010, Thebault2011, Giuppone2011}. Notably, the more recent studies by \cite{RafikovSilsbee2015a, RafikovSilsbee2015b} and \cite{SilsbeeRafikov2021} have led to the construction of comprehensive models that include gas drag, gravitational effects of an eccentric disk, and perturbations from an eccentric stellar companion, with the aim of investigating whether the fragmentation barrier can be overcome to allow for planetesimal growth. Further examples where both embryos and planetesimals were included in N-body simulations can be found in \cite{Lissauer2004}, \cite{Quintana2008}, and \cite{Quintana2007} \citep[with a comprehensive discussion in][]{Quintana2004}, as well as in the works of \cite{Thebault2004}, \cite{Turrini2004}, and \cite{TsukamotoMakino2007}. For the specific case of $\gamma$ Cephei, hydrodynamic simulations have been used to model the protoplanetary disk around both stars \citep{Nelson2000} and the gas accretion process onto the forming planet \citep{KleyNelson2008}, alongside investigations into habitability and dynamical stability \citep[e.g.,][]{Haghighipour2006}. Complementary to formation studies, dynamical analyses have aimed to identify stable regions for planetary survival in binary systems. Early works have used analytical approaches such as the Hill stability criteria \citep[e.g.,][]{Szebehely1980, MarchalBozis1982}, which rely on integrals of motion and zero-velocity surfaces to determine bounded orbits in simplified three-body setups. Other early efforts have turned to numerical integration in the three-body approximation \citep[e.g.,][]{Harrington1977, RablDvorak1988} to explore orbital stability in more general configurations. A landmark numerical study by \cite{HolmanWiegert1999} integrated the elliptic restricted three-body problem across a wide range of binary mass ratios and eccentricities, providing the widely used empirical formula for the outermost stable S-type orbit, $a_{\rm crit}$. Since then, the question of orbital stability has continued to be addressed, with the aim of extending the parameter space, also employing different tools. These include chaos indicators such as the fast Lyapunov indicator \citep[FLI;][]{Froeschle1997,Pilat-Lohinger2002, Dvorak2003, Dvorak2004}, Lyapunov exponents \citep{Benettin1980, Quarles2011}, the mean exponential growth factor of nearby orbits \citep[MEGNO;][]{Gozdziewski2001, Cincotta2003, Satyal2013} code, and frequency map analysis \citep[FMA;][]{Laskar1993, MarzariGallina2016}. Several studies directly solved Newton’s equations of motion with high-precision integrators, such as Bulirsh-Stoer, the Runge-Kutta-Fehlberg integration scheme or symplectic schemes \citep[e.g.,][]{David2003, musielak2005, Fatuzzo2006}, while others explored resonance overlap as a route to chaos \citep{MudrykWu2006} or related the onset of dynamical instability in hierarchical triple systems to chaotic energy and angular momentum exchange, analogously to tidally interacting binaries \citep{MardlingAarseth1999, MardlingAarseth2001}. Some other studies continued to use the concept of Jacobi’s integral and Jacobi’s constant to define the so-called surfaces of zero velocity \citep{Cuntz2007, Eberle2008}, while others introduced alternative definitions of stability based on geometric and topological properties, such as the hodograph eccentricity criterion \citep{EberleCuntz2010} or invariant loops \citep{Jaime2014}. Analytical and semi-analytical approaches were combined with dynamical maps to explore both short-term and long-term stability domains across the parameter space \citep{Andrade-InesMichtchenko2014}. More recent works \citep[e.g.,][]{Quarles2020} have mapped how the stability of S-type orbits depends on a broader set of binary and planetary orbital parameters, including inclinations and relative phases.

While these studies provide valuable insights into, for instance, the specific phases of planet formation or the dynamical stability of planetary orbits in binaries, they remain limited in scope. Formation models do not include disk evolution, while dynamical studies neglect realistic planet formation processes. As such, they are not embedded within a global, self-consistent framework where disk physics, accretion, migration, and stellar perturbations are treated simultaneously. A fully coupled model capable of evolving the protoplanetary disk and tracking planetary growth under the gravitational influence of a binary companion remains largely unexplored in the context of S-type binaries and it is the aim of this work. In \cite{PaperI}, we introduced our Planet formation Around bInaRy Stars (PAIRS) project, which is aimed at studying the global planet formation of S-type planets by adapting the Bern model. In Paper I, we discuss the modifications to the structure of the protoplanetary disk and how this modifies the planet formation picture compared to the single-star case. This companion work focuses on including the modifications to account for the gravitational interaction with the stellar companion and demonstrates how accounting for it can lead to additional differences from single-star systems, depending on binary parameters and the location of the planetary embryos with respect to the outer edge of the corresponding disk.

The paper is organized as follows. 
In Sect. \ref{sec:methods}, we present the project, focusing on the modifications to the N-body integrator that account for the gravitational influence of the stellar companion, along with a description of the setup of our planet formation simulations (Sect. \ref{subsec:initialconditions}).
In Sect. \ref{sec:results} we present our results, starting with our set A simulations (Sect. \ref{subsec:results_simulationsA}), which is aimed at evaluating the strength of the gravitational influence of the secondary star during in situ one-embryo planet formation simulations around the primary star. In \ref{subsec:results_simulationsB}, we address a smaller set of simulations with one initial planetary embryo (simulation set B) to show the difference in final planet properties when we compare S-type formation simulations to a single star case. In \ref{subsec:results_simulationsC}, we present the simulations with five initial embryos (simulation set C) comparing, once again, a S-type case to a single-star case. Finally, we discuss the results in Sect. \ref{sec:discussion} and present our conclusions in Sect. \ref{sec:conclusions}.

\section{Methods}\label{sec:methods}

Our PAIRS project was created to adapt the Bern model of planet formation and evolution \citep[see][for more detailed reviews]{Alibert2004,Alibert2005,Mordasini2012b,Mordasini2012a, Alibert2013,Fortier2013, NGPPS1, Brugger2018, Brugger2020, Voelkel2020} to the presence of a secondary star in a S-type configuration. The Bern model is a coupled formation and evolution model. First, it follows the main planet formation phase for 20 Myr, accounting for the evolution of the protoplanetary disk, formation and dynamics of pebbles, accretion of such pebbles and gas by the planetary embryos, planetary migration, and planet-planet interactions. After the main formation phase is over, the model tracks the thermodynamical evolution of each planet individually until 10 Gyr, accounting for processes such as atmospheric escape and tides. 

To include a stellar companion in an S-type configuration, we need to modify the gaseous disk prescription to incorporate disk truncation and tidal heating, then update the N-body integrator to account for gravitational interactions between the companion star and planetary bodies. Details of the disk-related modifications are presented in Paper I, along with a brief overview of the Bern model components left unchanged (see \cite{NGPPS1} for a more thorough description). In the following sections, we focus on the modifications to the N-body integrator, while noting additional Bern model components that were left unchanged, but not covered in Paper I.

\subsection{This work: Modifications to the N-body integrator}\label{sec:symplStype}

To account for the gravitational interaction between planetary bodies and secondary star, we modified the hybrid symplectic N-body integrator \textsc{Mercury}, implemented by \cite{Chambers99} and used in the Bern model to study planetary systems around single stars. We followed the theoretical prescription of \cite{Chambers02} for S-type systems, referred to as "wide binary integrator" by the authors. Here, we give a brief outline of the method and we refer to \cite{WisdomHolman} for a more detailed description
of the symplectic integrators. For the hybrid symplectic integrators we refer to \cite{Chambers99} and to \cite{Chambers02}, for the specific cases of S-type and P-type systems.

The symplectic integrator, first introduced by \cite{WisdomHolman}, is an optimal choice when we are seeking fast execution times and secular preservation of the energy. The main feature of this method is to split the Hamiltonian, $H$, of the system into two components (e.g., $H_A$ and $H_B$), which can be integrated analytically. Here, $H_A$ is the dominant operator and describes the Keplerian motion of the bodies in the system, while $H_B$ acts as a small perturbation and describes the gravitational interaction among bodies. The problem with this original method is that it does not include the possibility of having close encounters. When two bodies get very close to each other, the respective interaction term inside the Hamiltonian, $H_B$, can become greater than the Keplerian Hamiltonian, $H_A$, breaking down the fundamental condition $\abs{H_B}\ll\abs{H_A}$ of symplectic integrators \citep{Chambers99}. To solve this issue, \cite{Chambers99} introduced a hybrid symplectic algorithm, namely: keeping the same Hamiltonian splitting of \cite{WisdomHolman}, when a close encounter takes place, the interaction term that becomes dominant can be moved to the Keplerian Hamiltonian. It then has to be integrated numerically with a conventional, high-precision integrator for that specific time step (e.g., Bulirsh-Stoer integrator). While this hybrid symplectic method is designed for single-star systems, \cite{Chambers02} offered a theoretical prescription for studying binary systems, while allowing for close encounters to be handled as well.

The Hamiltonian for a system with two stars and N planets in an S-type configuration, expressed in inertial coordinates (\vec{x},\vec{p}), is given by Eq. (8) of \cite{Chambers02}. The new set of coordinates, \vec{X}, and conjugate momenta, \vec{P,} (henceforth referred to as the S-type coordinates) are defined by Eqs. (9) and (10) of \cite{Chambers02}. They allow us to split the system's Hamiltonian into three components, $H_{\text{kep}}$, $H_{\text{int}}$, and $H_{\text{jump}}$, identified by Eq. (12) of \cite{Chambers02}. In this case, $H_{\text{kep}}$ is usually the dominant term and describes the Keplerian motion of the secondary star and the planets, $H_{\text{int}}$ contains the planet-interaction terms and terms related to gravitational influence of the stellar companion on the planets, and $H_{\text{jump}}$, contains terms that arise from the kinetic energy of the star that hosts the planets. To integrate each one of these terms, the integration step is divided into five sub-steps \citep{Chambers02}:
\begin{itemize}
     \item Advance $H_{\text{int}}$ for $\tau/2;$
     \item Advance $H_{\text{jump}}$ for $\tau/2;$
     \item Advance $H_{\text{kep}}$ for $\tau;$
     \item Advance $H_{\text{jump}}$ for $\tau/2;$
     \item Advance $H_{\text{int}}$ for $\tau/2.$
\end{itemize}

The momenta, $\vec{P}$, are related to the pseudo-velocities, $\vec{V,}$ rather than the real velocities, $\dot{\vec{X}}$. \citet{Verrier} provided their relation, which we report in Appendix \ref{sec:coordinate_trasformations} (see Eqs. \eqref{P&pseudoV}--\eqref{V&pseudoV}). In Appendix \ref{sec:coordinate_trasformations} we also provide the coordinate transformations from S-type to heliocentric coordinates (see Eqs. \eqref{Xtohelio}--\eqref{heliotoV}).

The evolution of the interaction Hamiltonian modifies the pseudo-velocities $\vec{V}$ of the secondary star and planets by quantities proportional to the respective acceleration terms. In particular, Eqs. (14) and (15) of \citet{Chambers02} define the $x$ component of the acceleration of planet $k$ and of the secondary star. On the other hand, $H_{\text{jump}}$ modifies only the positions $\vec{X}$ \citep{Chambers02} and the evolution of this part of the Hamiltonian is
\begin{equation}
     \dv{X_i}{t} = \sum_j \dfrac{P_j}{m_A}.
\end{equation}

When close encounters take place, the condition $|H_{\text{kep}}|\gg|H_{\text{int}}|$ breaks down because the interaction term among the planets that undergo a close encounter becomes dominant. The solution from \citet{Chambers99} proposes moving this term to $H_{\text{kep}}$ by introducing the function $K(r_{ij})$ and redistributing the terms in the Hamiltonian as in Eq. (9) of \citet{Chambers99} (or Eq. (7) of \cite{Chambers02}). Details on how to choose the function $K(r_{ij})$ are described in Sects. 4.1 and 4.2 of \cite{Chambers99}. However, compared to the function $K$ defined in Eq. (10) of \citet{Chambers99}, whose derivative displays discontinuities in $y=0$ and $y=1$, a more suitable equation that would avoid such a problem is the one used in the \textsc{Mercury} package \citep{Chambers99}, expressed as
\begin{equation}\label{Kfunc}
     K=
     \begin{cases}
          0\quad & y<0,\\
          10y^3-15y^4+6y^5\quad & 0<y<1,\\
          1\quad & y>1,
     \end{cases}
\end{equation}
with $y=\dfrac{R_{ij}-0.1r_{\text{crit}}}{0.9r_{\text{crit}}}$, where $R_{ij}$ is the distance between the two bodies having the close encounter and $r_{\text{crit}}$ determines when an encounter between two bodies takes place \citep{Chambers99}. With this prescription, the Keplerian Hamiltonian can not be solved analytically and, in our simulations, the integration during a close encounter is done with the Bulirsh-Stoer integrator, which is implemented in the \textsc{Mercury} package \citep{Chambers99}. 

We implemented this symplectic mapping to work alongside the \textsc{Mercury} version included in the Bern Model of planet formation. To do so, we followed \citet{TurriniCodice}, who implemented this mapping in the same framework of the \textsc{Mercury} code and we tested our implementation by comparing the outcome of our simulations to the output of simulations, with same initial conditions, run with the Dynamical Plug-In (DPI) library by \cite{TurriniCodice}. We present such a comparison in Appendix \ref{sec:Nbody_testing}. We also assessed the accuracy of our integrator for different binary architectures and initial planet locations through pure three body simulations, which we present in Appendix \ref{sec:set0}. Finally, we note that within the Bern model, the N-body integrator is kept running for 20 million years.

\subsection{Collision detection and treatment}

 Although we modified the structure of the N-body integrator to account for the gravitational influence of the stellar companion, we retained the original approach for detecting and handling planet-planet collisions \citep{NGPPS1}. A collision occurs when the mutual distance between two planets falls below the sum of their radii. However, during the attached phase, the planet’s radius must be approximated, since the N-body integrator does not have access to the full envelope structure. In this phase, the collision detection radius is computed by assuming that the total planetary mass has the same density as its core. Once a planet transitions to the detached phase, the collision detection radius is instead given by the planetesimal capture radius, following \citet{InabaIkoma2003}. When a collision is detected, the cores of the two planets merge, while the envelope of the less massive body is ejected. The energy from the impact is then added as an additional luminosity contribution in the structure calculation of the resulting planet \citep{NGPPS1}.

\subsection{Orbital migration and additional forces}

For Type I migration, the model employs the approach of \cite{ColemanNelson2014}, which computes the torques as in \cite{Paardekooper2011} including the reduction of the corotation torque due to eccentricity and inclination as in \cite{BitschKley2010}. For Type II migration, we followed the nonequilibrium approach of \cite{Dittkrist2014} and the switch from Type I to Type II migration can be computed with the gap opening criterion of \cite{Crida2006}.
Migration and damping of eccentricity and inclination are included in the N-body as additional force terms. We refer to Eqs. (104), (105), and (106) of \cite{NGPPS1} for further details.

\subsection{Initial conditions}\label{subsec:initialconditions}

The planetary bodies are assumed to be orbiting the more massive star of the binary (i.e., the primary star), while we refer to the less massive stellar companion as the secondary star. We analyzed S-type planet formation from three different perspectives: parameter space exploration (simulation set A), single-versus-binary star for one-embryo simulations (simulation set B), and multi-embryo configurations (simulation set C). We also performed in situ planet formation simulations with a similar setup to set A, but modeling the formation around the less massive, secondary star. We present these simulations in Appendix \ref{sec:circumsecondary}. 

\subsubsection{Simulation set A: Parameter space exploration}

To assess the impact of the secondary star’s gravitational influence on growing planetary bodies and quantify its strength across different binary configurations, we performed a grid of 5000 simulations with initial conditions outlined in the left column of Table \ref{tab:initialconditionsABC}. We fixed the primary star's mass $M_1$ at 1 \Msun\ and randomly set the secondary star's mass $M_2$ between 0.1 and 1 \Msun\ to sample different binary mass ratios $q=M_2/M_1$. The binary separation, \abin\,, and eccentricity, \ebin\,, were randomly sampled within the ranges 10-1000 au and 0-0.9, respectively. In each simulation, we placed one Moon-mass embryo (i.e., $M_\text{p}=10^{-2}$ \Mearth), with an initial location, $a_{\text{p,0}}$, randomly drawn between 1 au and the location of the truncation radius, $R_{\rm trunc}$, minus 1 au. All these parameters were sampled with a uniform or log-uniform distribution, as reported in Table \ref{tab:initialconditionsABC}. Here, $R_{\rm trunc}$ is defined by \cite{Manara2019} as $R_{\rm trunc} (M_1, M_2, e_{\rm bin}, a_{\rm bin})= R_{\rm Egg}\times (b e_{\rm bin}^c +h \mu^k).$ We refer to Paper I and to \cite{Manara2019} for further details. Regarding the parameters for the protoplanetary disk, we built a sampling using a log-uniform distribution of initial gas mass before truncation $M_{\text{gas, b.t.}}$ in the range [0.001,0.1] \Msun, dust-to-gas ratio in the range [0.0056,0.03162], disk viscosity parameter $\alpha$ \citep{ShakuraSunyaev1973} in the range [0.0001,0.001], and a disk characteristic radius, $r_{\rm char}$, in the range [10,200] au. These ranges of parameters are physically motivated by observations of single-star disks and were used also in the works of \cite{NGPPS_II} and \cite{Weder2023}. The left column of Table \ref{tab:initialconditionsABC} summarizes all these initial conditions along with the planetary envelope opacity, $\kappa$, which we kept fixed at 0.01 of the opacity of the interstellar medium, justified by past works \cite[e.g.,][]{MovshovitzPodolak2008}. We note that this grid is the same as the one presented in Paper I, with the difference that we have included the gravitational interaction between secondary star and planetary bodies.

\begin{table*} [t]
    \caption{Initial conditions for our simulation set A (left column), B (middle column), and C (right column).}
    \label{tab:initialconditionsABC}
    \centering
    \begin{tabular}{l|l|l|l}
        \hline
        \hline
        & Simulation set A & Simulation set B & Simulation set C \\
        \hline
        \multicolumn{4}{c}{disk parameters}\\
        \hline
          $M_{\text{gas, b.t.}}$ [\Msun] & Log$\mathcal{U}$[0.001,0.1] & 0.1 & 0.1\\
          dust-to-gas ratio & Log$\mathcal{U}$[0.0056,0.03162] & 0.01 & 0.01 \\
          $\alpha$ & Log$\mathcal{U}$[0.0001,0.001] & 0.001 & 0.001\\
          $r_{\rm char}$ [au] & Log$\mathcal{U}$[10,200] & 50 & 50 \\
          \hline
        \multicolumn{4}{c}{Binary parameters}\\
        \hline
          $M_1$ [\Msun] & 1 & 1 & 1\\
          $M_2$ [\Msun] & $\mathcal{U}$[0.1, 1] & 0.5 & 0.5\\
          \ebin & $\mathcal{U}$[0, 0.9] & 0 & 0\\
          \abin\ [au] & Log$\mathcal{U}$[10, 1000] & 20, 50, 75, 100, 300 & 100\\
        \hline
        \multicolumn{4}{c}{Planet parameters}\\
        \hline
          $N_{\text{p}}$ & 1 & 1 & 5 \\
          $a_{\text{p,0}}$ & Log$\mathcal{U}$[1 au, $R_{\text{trunc}}$ - 1 au] & 5, 20 [au] & 2, 4, 8, 10, 20 [au] \\
        $\kappa$ & 0.01$\times$BL94 & 0.01$\times$BL94 & 0.01$\times$BL94\\
         \hline
    \end{tabular}
    \tablefoot{BL94: \citet{BellLin1994}, $M_{\text{gas, b.t.}}$ is the mass of gas before truncation.}
\end{table*}

\subsubsection{Simulation set B: Binary versus single star in a one-embryo setup}

Simulation set B focuses on a smaller subset of planet formation simulations, with the aim of comparing them to single-star cases to highlight the differences in the final planet's properties. We simulated circular binaries with a 1 \Msun\ primary and a 0.5 \Msun\ secondary, exploring four binary separations: 20, 50, 75, and 100 au. In each simulation, we placed one (i.e., $N_{\text{p}}$ = 1) Moon-mass embryo at initially 5 au or 20 au from the primary star. In the 20 au case, we discarded the simulations with \abin\ = 20, 50 au because the corresponding disks were truncated at less than the initial location of the embryo (i.e., the embryo would end up outside of the disk, see Table \ref{tab:resultsB}). For this reason, in the case where the embryo is at 20 au, we introduced one additional binary simulation with \abin\ = 300 au. In the middle column of Table \ref{tab:initialconditionsABC}, we summarize these initial conditions and those of the protoplanetary disk, which we kept fixed to a single value within the range used in set A. Furthermore, each one of these simulations is performed twice, once assuming in situ formation and the other allowing the planet to migrate within the disk.

\subsubsection{Simulation set C: Binary versus single star in a multi-embryo setup}
For our multi-embryo simulations (set C), we compared a circular binary with \abin\ = 100 au to a single-star system. In both scenarios, five Moon-mass embryos (i.e., $N_{\text{p}}$ = 5) were initially positioned at corresponding radial distances, $a_{\text{p,0}}$, from the central star ($M_1$ = 1 \Msun), ensuring comparability between the two cases. In the right column of Table \ref{tab:initialconditionsABC} we summarize the initial conditions for this set of simulations. 

\subsection{Metrics}
To quantify the strength of the gravitational influence of the secondary star on planetary bodies, we employed two metrics, $\mathcal{M}_1$ and $\mathcal{M}_2$. Here, $\mathcal{M}_1$ identifies how close the planet is to the location of the truncation radius, where the tidal interaction of the secondary star and on disk is the strongest, while $\mathcal{M}_2$ quantifies how close the planet is to the edge of the Hill sphere of influence of the primary star. We define these metrics as
\begin{equation}\label{metric1&2}
        \mathcal{M}_1 = \dfrac{a_\text{p,0}}{R_\text{trunc}}\quad ; \quad \mathcal{M}_2 = \dfrac{a_\text{p,0}}{R_\text{Hill,1}}
,\end{equation}
where $a_\text{p,0}$ is the initial location of the planet, $R_\text{trunc}$ is the disk truncation radius, and $R_\text{Hill,1}$ is the radius of the Hill sphere of the primary star. The Hill radius defines the regions in which the gravity of the primary star dominates over the gravity of the secondary star, and it is defined as

\begin{equation}\label{Hill_sphere}
    R_\text{Hill,1} = a_\text{bin}\cdot(1-e_\text{bin})\cdot\left[\dfrac{M_1}{3\cdot(M_1 + M_2)}\right]^{1/3}
.\end{equation}

Typically, planets that orbit within 1/2 of $R_\text{Hill,1}$ are stable, as the primary star's gravitational influence outweighs the perturbations coming from the stellar companion. However, some planets near $ \sim R_\text{Hill,1}/2$ could experience significant secular perturbations from the secondary star and become, for example, ejected from the system. Once a planet’s orbit extends beyond 1/2 of $R_\text{Hill,1}$, it becomes unstable due to strong tidal forces from the secondary star, leading to high eccentricity growth and ejections \citep[see, e.g., the study of satellite stability from][]{Domingos2006}.

Many studies have addressed the problem of stability in the case of S-type systems and some have given analytical or semi-analytical expressions for the critical stability boundary, which identifies the maximum semimajor axis that an S-type orbit can have to remain stable \citep{HolmanWiegert1999}. In our analysis, alongside the metrics defined above, we compared our results to the critical stability limit expression obtained by \cite{Quarles2020}, defined as

\begin{equation}\label{acrit}
    \begin{split}
        a_\text{crit}/a_\text{bin} = & (0.501\pm0.002) - (0.435\pm0.003)\mu + \\
        & - (0.668\pm0.009)e_\text{bin} + (0.152\pm0.011)e_\text{bin}^2 + \\
        & + (0.644\pm0.015)\mu e_\text{bin} - (0.196\pm0.019)\mu e_\text{bin}^2
    \end{split}
.\end{equation}
Here, $\mu$ is the binary mass parameter and it is equal to $M_2/(M_1+M_2)$ if we are modeling the formation around the primary star; whereas it becomes $M_1/(M_1+M_2)$ if we are modeling the formation around the secondary. If the planet semimajor axis is greater than $a_\text{crit}$, then it is more likely to become unstable. 

\section{Results}\label{sec:results}

\subsection{Simulation set A: Parameter space exploration}\label{subsec:results_simulationsA}

\begin{figure}[h!]
        \centering
        \includegraphics[width=0.9\hsize]{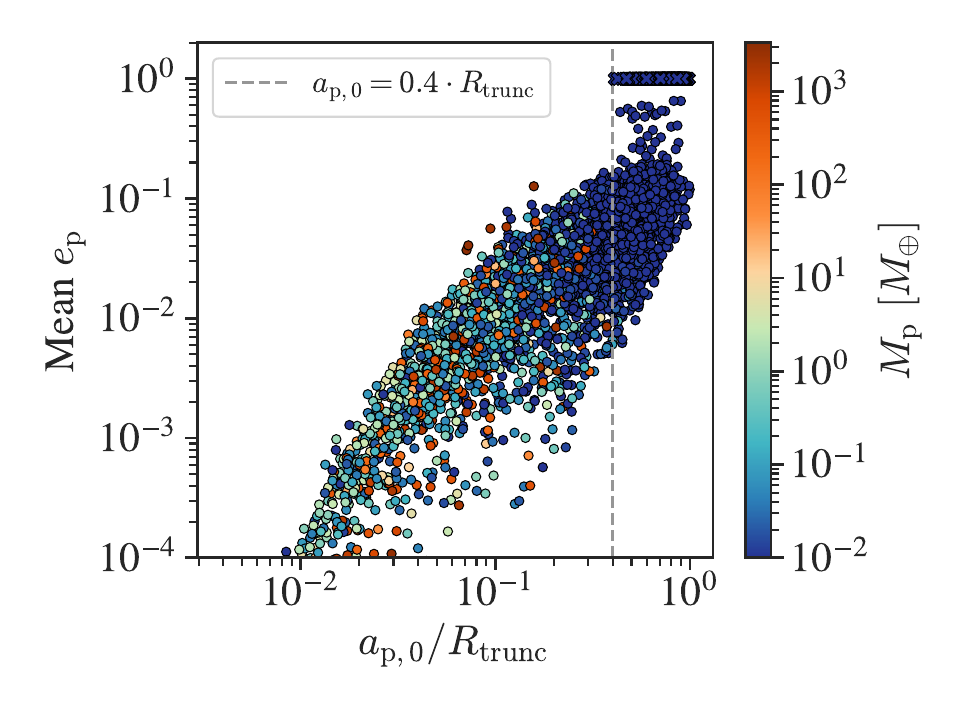}
        \includegraphics[width=0.9\hsize]{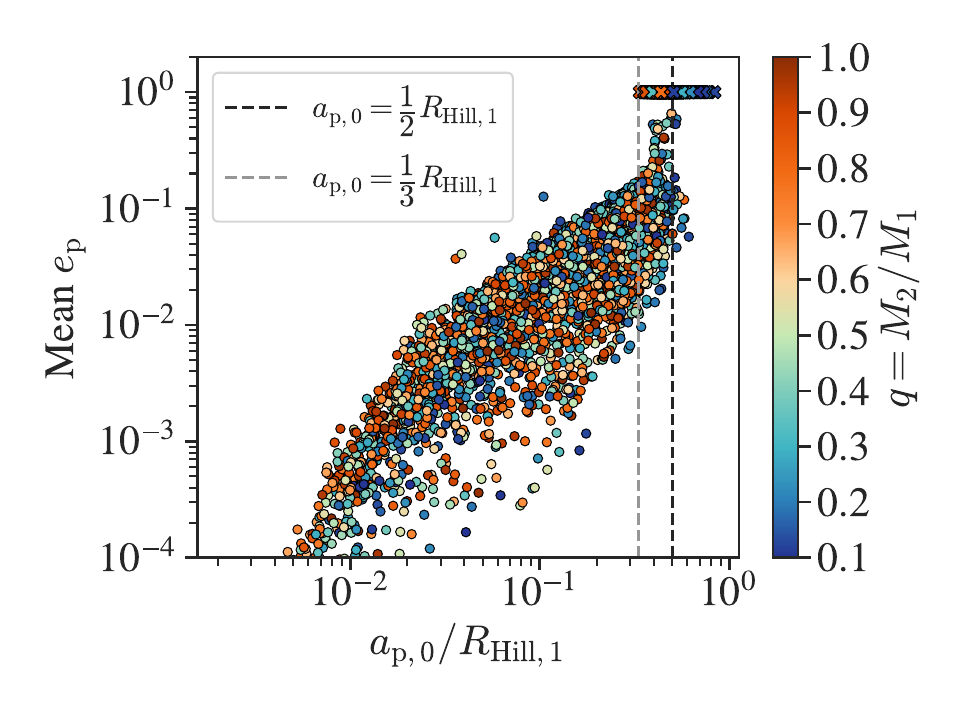}
        \vspace{-0.5cm}
        \caption{Simulation set A: Initial planet location relative to the disk truncation radius (top: $a_{\text{p,0}}/R_\text{trunc}$), and to the primary star’s Hill radius (bottom:\ $a_{\text{p,0}}/R_\text{Hill,1}$). Top panel: Color bar shows the final planet mass and the gray dashed line marks $0.4\cdot R_\text{trunc}$. Bottom panel: Color bar shows the binary mass ratio and the gray and black dashed lines mark $1/3$ and $1/2\cdot R_\text{Hill,1}$. Circles denote surviving planets, while crosses indicate planets lost by ejection or instability from close encounters with the central star.}
        \label{insitu_grid1}
\end{figure}

In set A, we simulated a grid of 5000 systems with varying disk parameters ($M_{\rm gas, b.t.}$, dust-to-gas ratio, $\alpha$, $r_{\rm char}$), binary parameters ($q$, \abin, \ebin) and different initial embryo locations $a_\text{p,0}$, as outlined in the left column of Table \ref{tab:initialconditionsABC}. In our simulations a body can be lost from the system through an ejection (i.e., if its orbital distance exceeds the outer boundary of the disk), but also if it has a close encounter with the central star. When either of these two events took place, we removed these bodies from our simulations and we set their eccentricity to unity. Out of our simulations, 93.2\% systems retain their planets, while 3.1\% systems lose their planets due to ejection and 3.7\% systems become unstable due to a close encounter between the planet and the central star. Figure \ref{insitu_grid1} shows the mean planet eccentricity as a function of $\mathcal{M}_1 = a_{\text{p,0}}/R_\text{trunc}$ (top) and as function of $\mathcal{M}_2 = a_{\text{p,0}}/R_\text{Hill,1}$ (bottom). Circle markers represent systems that retain their planets, while cross markers indicate planets lost due to ejection or instabilities due to a close encounter with the central star, respectively. The color-coding in the top and bottom panels reflect the final planet mass and the binary mass ratio, respectively. We find that planets initially located beyond 0.4 of the disk truncation radius ($\mathcal{M}_1>0.4$, to the right of the gray dashed line in the top panel of Figure \ref{insitu_grid1}) start to be lost. This trend is directly linked to the planet’s location relative to the primary star’s Hill radius, as shown in the bottom panel of the same figure. When $\mathcal{M}_2>1/3$, planet loss begins to occur, and its frequency increases as planets approach 1/2 of $R_\text{Hill,1}$.
Specifically, 28.6\% of systems with $1/3<\mathcal{M}_2<1/2$ and 86.5\% of systems with $\mathcal{M}_2>1/2$ experience the loss of their planet. Furthermore, planets that are ultimately lost never accreted material: dynamical instabilities act early, before pebble accretion begins, leaving them at their initial mass of $\sim 10^{-2}$ \Mearth\ (color-coding in the top panel of Figure \ref{insitu_grid1}). 

\begin{figure}
        \centering
        \includegraphics[width=0.9\hsize]{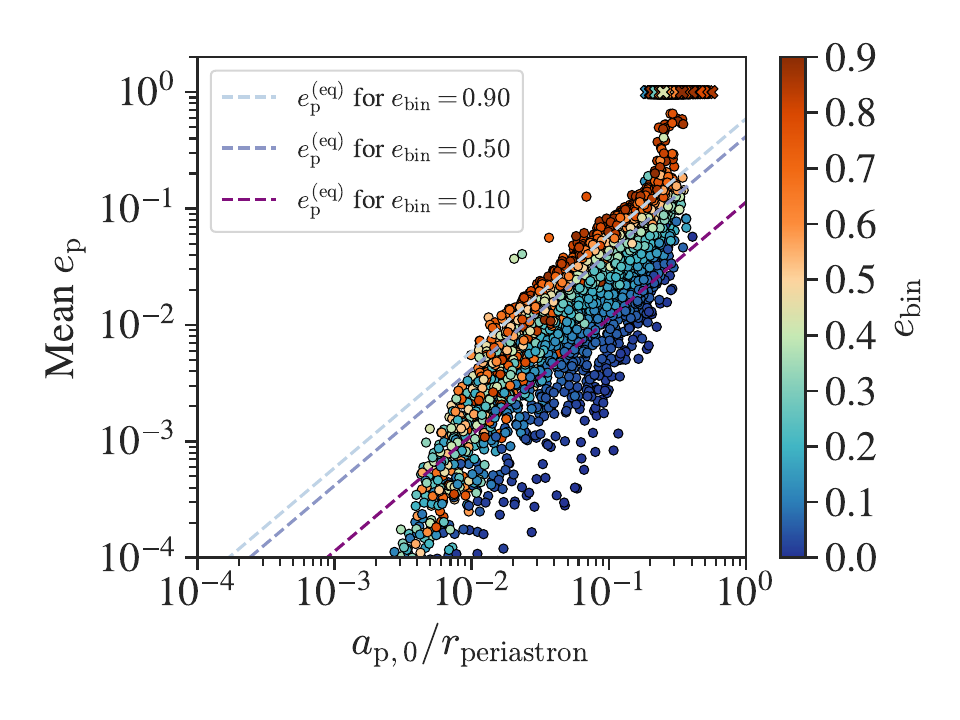}
        \vspace{-0.5cm}
        \caption{Simulation set A: Initial planet location relative to the binary periastron distance ($a_{\text{p,0}}/r_\text{periastron}$) versus mean planet eccentricity. The color bar shows the binary eccentricity and purple dashed lines indicate the equilibrium eccentricity, $e_\text{p}^\text{(eq)}$, \citep[Eq. (14)][]{Mardling2007} for three binary eccentricities. Circles denote surviving planets, while crosses indicate planets lost by ejection or instability from close encounters with the central star.}
        \label{insitu_grid2}
\end{figure}

 From the color-coding in the bottom panel of Figure \ref{insitu_grid1}, we did not find any clear correlation between planetary eccentricity and binary mass ratio, there is a correlation with the binary eccentricity. This is shown in Figure \ref{insitu_grid2}, where we plot the mean planet eccentricity as a function of the planet location relative to the binary periastron distance, $a_{\text{p,0}}/r_\text{periastron}$. Stable planets are represented as circles, while lost planets are shown with crosses. From the color-coding, which represents the binary eccentricity, we find that higher binary eccentricity results in higher planetary eccentricity. This correlation is also responsible for the slope of the curve both Figure \ref{insitu_grid1} and \ref{insitu_grid2}. Indeed, \cite{Mardling2007} derived an expression for the planet equilibrium eccentricity, $e_\text{p}^\text{(eq)}$ defined in their Eq. (14), which we display, for three different values of the binary eccentricity, as purple dashed lines in the same Figure \ref{insitu_grid2}. Furthermore, we find that 58.9\% of lost planets belonged to a highly eccentric binary (\ebin\ > 0.5).

\begin{figure}
        \centering
        \includegraphics[width=0.9\hsize]{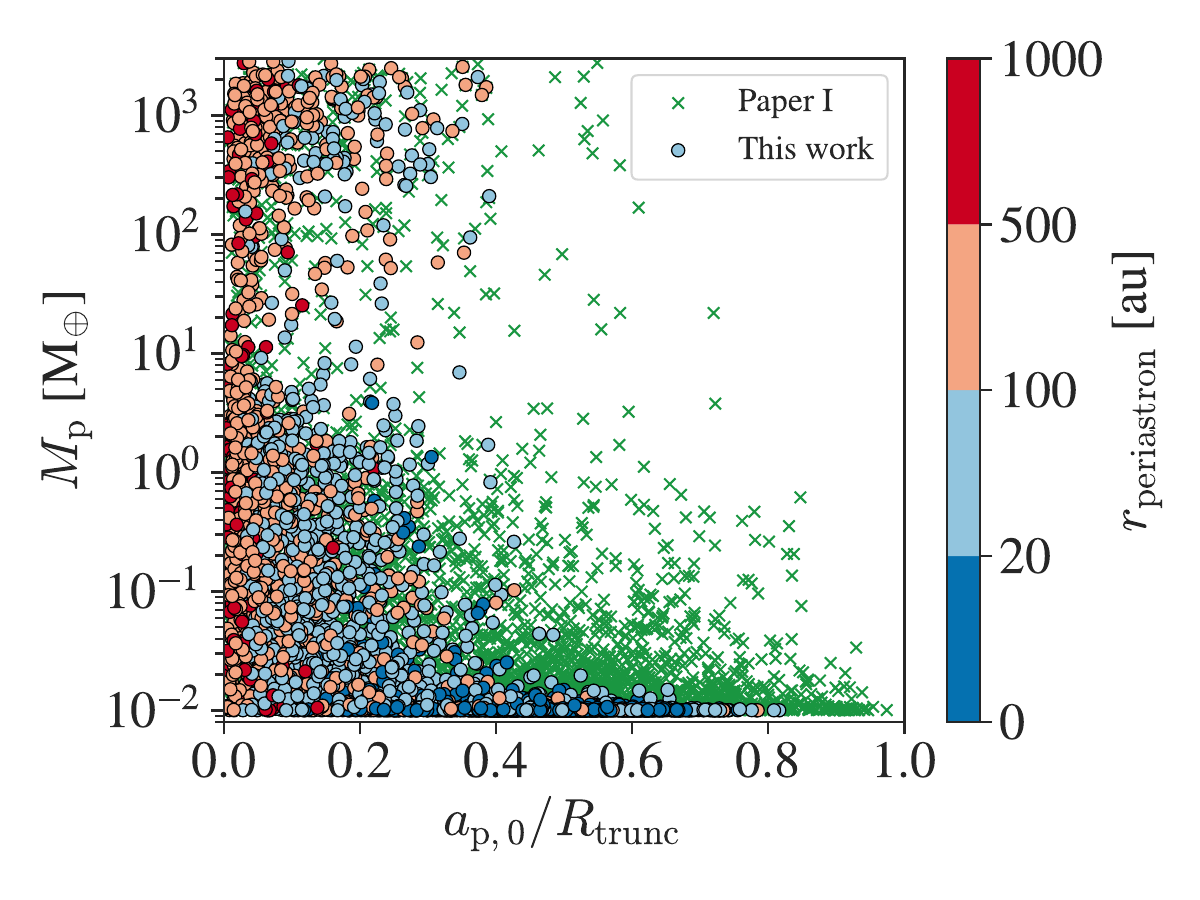}
        \vspace{-0.5cm}
        \caption{Simulation set A: Initial planetary position relative to the disk truncation radius, $a_{\text{p,0}}/R_\text{trunc}$, versus the final planet mass. The color bar shows the binary periastron distance. Circle markers represent stable planets from our simulations (effect of disk truncation + gravitational perturbation inside the N-body integrator), while green crosses represent stable planets from Paper I simulations (effect of disk truncation only).}
        \label{insitu_grid3}
\end{figure}

Planets that survive beyond $\mathcal{M}_1=0.4$ are also inefficient at growing. This is shown by the color-coding of the top panel of Figure \ref{insitu_grid1} and by Figure \ref{insitu_grid3}, which displays the final planet mass as a function of $\mathcal{M}_1$, color-coded with the binary periastron distance. Stable planets from our simulations are shown as circles, while stable planets from the Paper I simulations are shown as green crosses. We find that if $\mathcal{M}_1\lesssim$ 0.4, and especially if $\mathcal{M}_1$ < 0.3, planets are able to grow up to $\sim 6$ \Mj, but only if hosted by binaries with $r_\text{periastron}$ > 20 au. The majority of these giant planets anyways forms within binaries that have $r_\text{periastron}$ > 100 au. However, as the embryo is placed further out in the disk (i.e., $\mathcal{M}_1$) increases, the final planet mass decreases sharply if we include the gravitational interaction together with the effect of disk truncation. To quantify this effect, we calculated the mean planet mass in the bins $\mathcal{M}_1 < 0.3$, $0.3<\mathcal{M}_1 < 0.4$, $0.4<\mathcal{M}_1 < 0.5$, and $\mathcal{M}_1 > 0.5$,  finding that in the first bin the mean mass is $\sim$ 112 \Mearth\,, both for this work and Paper I. Already at the second bin, $0.3<\mathcal{M}_1 < 0.4$ the mean mass starts to decrease; namely, we obtained a mean mass of $\sim$26 \Mearth\ compared to a mean mass of $\sim$44 \Mearth\ from Paper I, which corresponds to a reduction in mass of about 41\% when including the gravitational influence of the companion. Furthermore, in the range $0.3 \lesssim \mathcal{M}_1 < 0.4$, the inclusion of binary perturbations reduces the number of planets above 0.1 \Mearth\ by $\sim$79\%. From the third bin onward, which corresponds to $\mathcal{M}_1 \gtrsim 0.4 $, we start to have even more pronounced differences. In our simulations we find that from this region onward the mean planet mass is $\sim 0.01$ \Mearth, while in Paper I we can still form planets with at least $\sim 0.1$ \Mearth\ until $\mathcal{M}_1 \sim$0.8. This behavior is partly due to the rapid inward drift of pebbles, which quickly depletes the outer disk (Paper I), but also to the additional gravitational perturbations from the stellar companion, which excite the planetary eccentricities and reduce the accretion efficiency. 

\begin{figure}
        \centering
        \includegraphics[width=0.9\hsize]{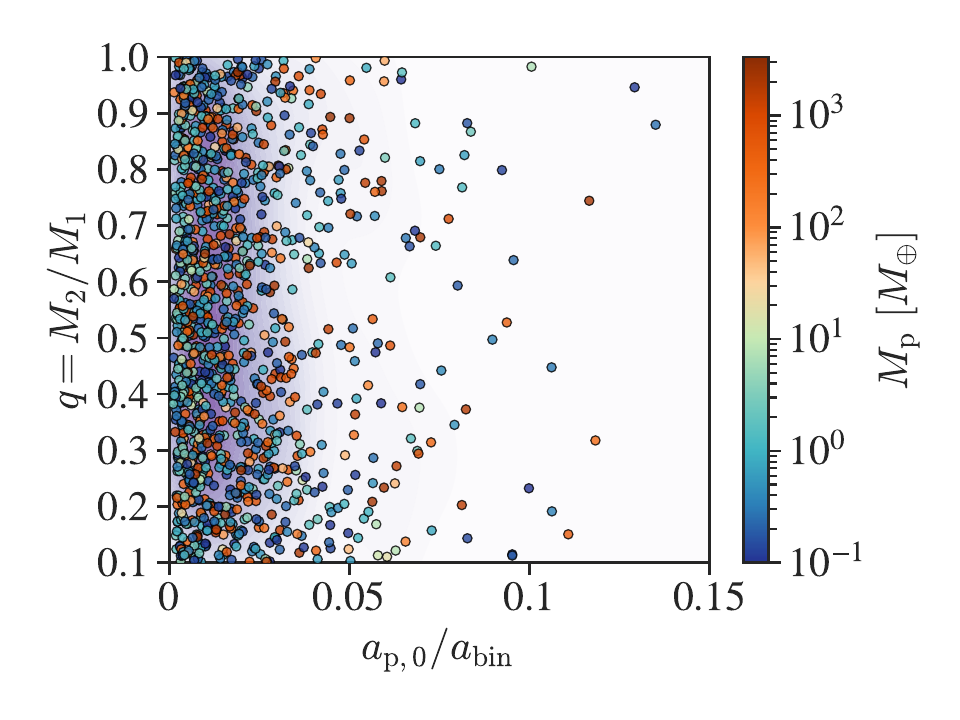}
        \includegraphics[width=0.9\hsize]{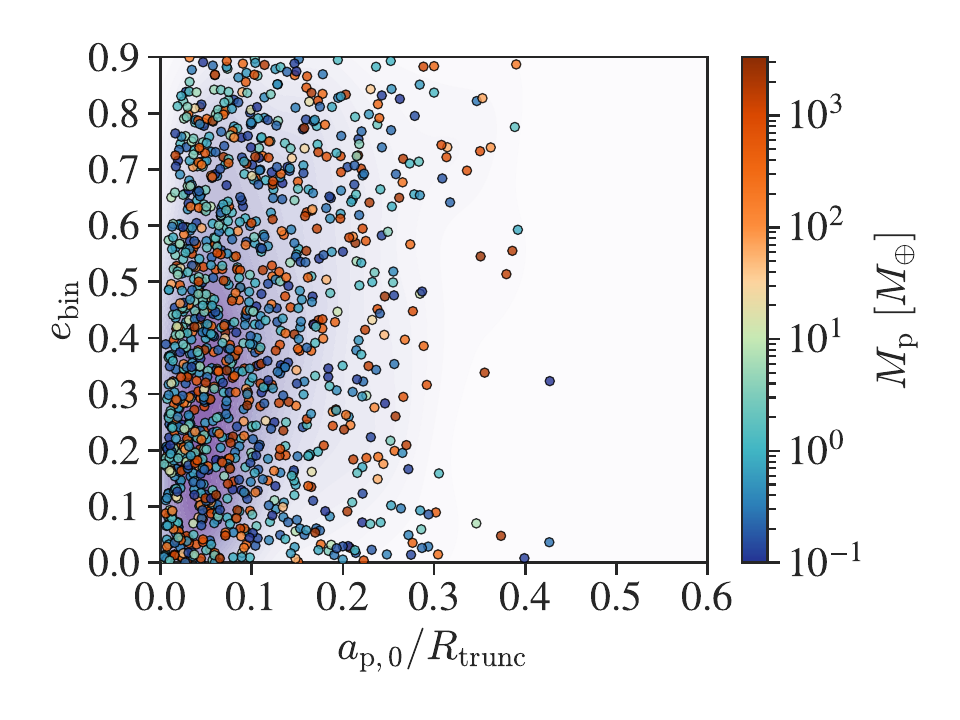}
        \vspace{-0.5cm}
        \caption{Simulation set A: Initial planetary position relative to the binary separation, $a_{\text{p,0}}/a_\text{bin}$, versus the binary mass ratio (top panel) and initial planetary position relative to the disk truncation radius, $a_{\text{p,0}}/R_\text{trunc}$, versus the binary eccentricity (bottom panel). We display planets that became at least more massive than Mars and their color reflects their final planet mass. The shaded contours highlight regions of highest planet density, obtained through a two-dimensional kernel density estimation (KDE) considering planets with masses above 10 \Mearth.}
        \label{insitu_grid4}
\end{figure}

Looking at the systems which formed planets more massive than Mars, the top panel of Figure \ref{insitu_grid4} shows the binary mass ratio as a function of initial planet location relative to the binary separation, $a_{\text{p,0}}/a_\text{bin}$. The bottom panel of the same figure shows the binary eccentricity versus $\mathcal{M}_1$. In both panels, the color-coding reflects the final planet mass and the purple contour region highlights the highest density of planets above 10 \Mearth. Planets more massive than Mars form only within 0.15 \abin, with most of them located inside 0.05 \abin. In terms of disk size, these planets form within 0.5 $R_{\rm trunc}$, with the majority located inside 0.3 $R_{\rm trunc}$. In this inner region, some planets grow as large as $\sim 6$ \Mj, independently of both the binary mass ratio and binary eccentricity. However, growth does indeed depend on the binary periastron distance, as we show in Figure \ref{insitu_grid3}. 

Lastly, we compared our results to the stability limit $a_{\text{crit}}$ of \cite{Quarles2020}, defined in Eq. \eqref{acrit}. We find that 78.4\% of planets located beyond this limit become unbound, compared to only 3.3\% of those located within the stable region.

\begin{table}[h]
    \caption{Simulation set B: Binary separation, \abin, truncation radius, $R_\text{trunc}$, critical semimajor axis, $a_{\text{crit}}$, initial planet location, $a_{\text{p,0}}$, and metrics, $\mathcal{M}_{1,2}$, for the different simulated scenarios.}
    \centering
    \begin{tabular}{l|l|l|l|l|l}
        \hline
        \hline
          \abin\  [au] & $R_\text{trunc}$ [au] & $a_{\text{crit}}$ [au] & $a_{\text{p,0}}$ [au] & $\mathcal{M}_1$ & $\mathcal{M}_2$\\
           \hline
           20 & 7.66 & 7.12 & 5 & 0.65 & 0.41\\
           \hline
           50 & 19.15 & 17.80 & 5 & 0.26 & 0.17\\
           \hline
           75 & 28.72 & 26.70 & 5 & 0.17 & 0.11\\
           75 & 28.72 & 26.70 & 20 & 0.69 & 0.44\\
           \hline
           100 & 38.30 & 35.60 & 5 & 0.16 & 0.08\\
           100 & 38.30 & 35.60 & 20 & 0.65 & 0.33\\
           \hline
           300 & 114.89 & 106.80 & 20 & 0.17 & 0.11\\
         \hline
    \end{tabular}
    \label{tab:resultsB}
\end{table}

\begin{figure*}[b]
  \centering
  \begin{minipage}[t]{0.74\textwidth}
    \centering
    \begin{minipage}[t]{0.495\linewidth}
      \centering
      \includegraphics[width=\linewidth]{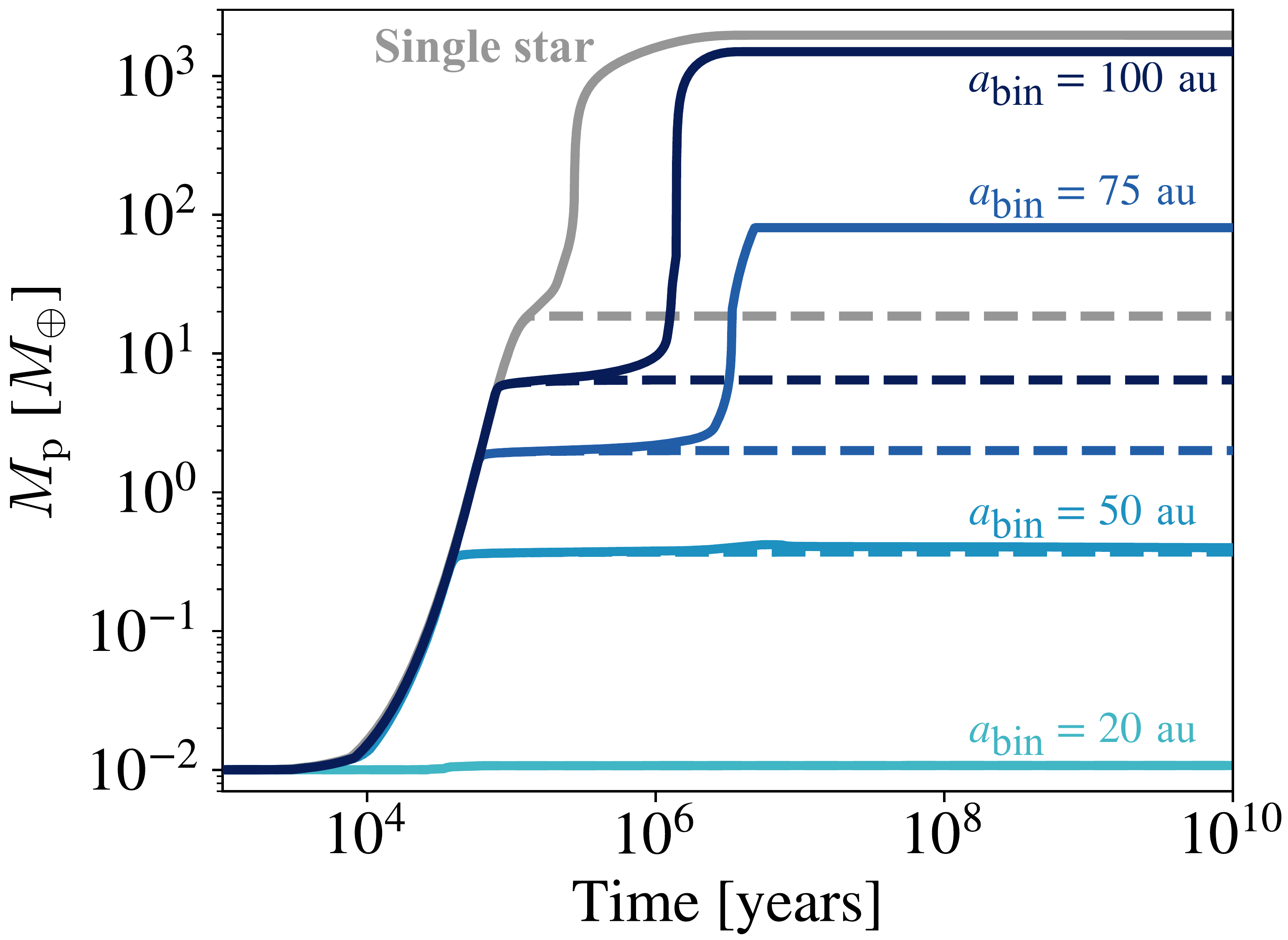}
      
      {\small (a) Mass evolution: embryo at 5 au}
    \end{minipage}
    \hfill
    \begin{minipage}[t]{0.495\linewidth}
      \centering
      \includegraphics[width=\linewidth]{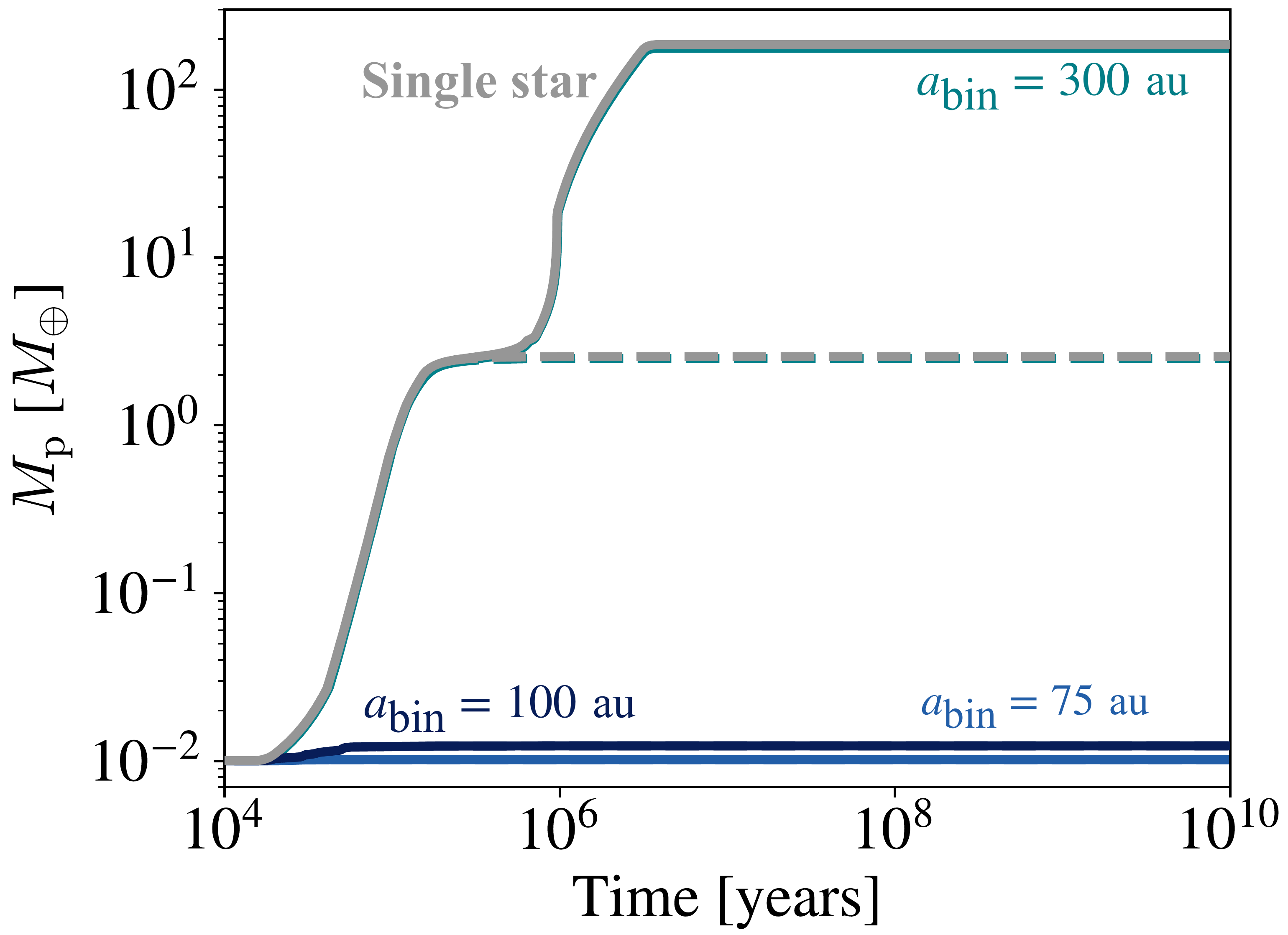}
      
      {\small (b) Mass evolution: embryo at 20 au}
    \end{minipage}
    \\[3mm]
    \begin{minipage}[t]{0.495\linewidth}
      \centering
      \includegraphics[width=\linewidth]{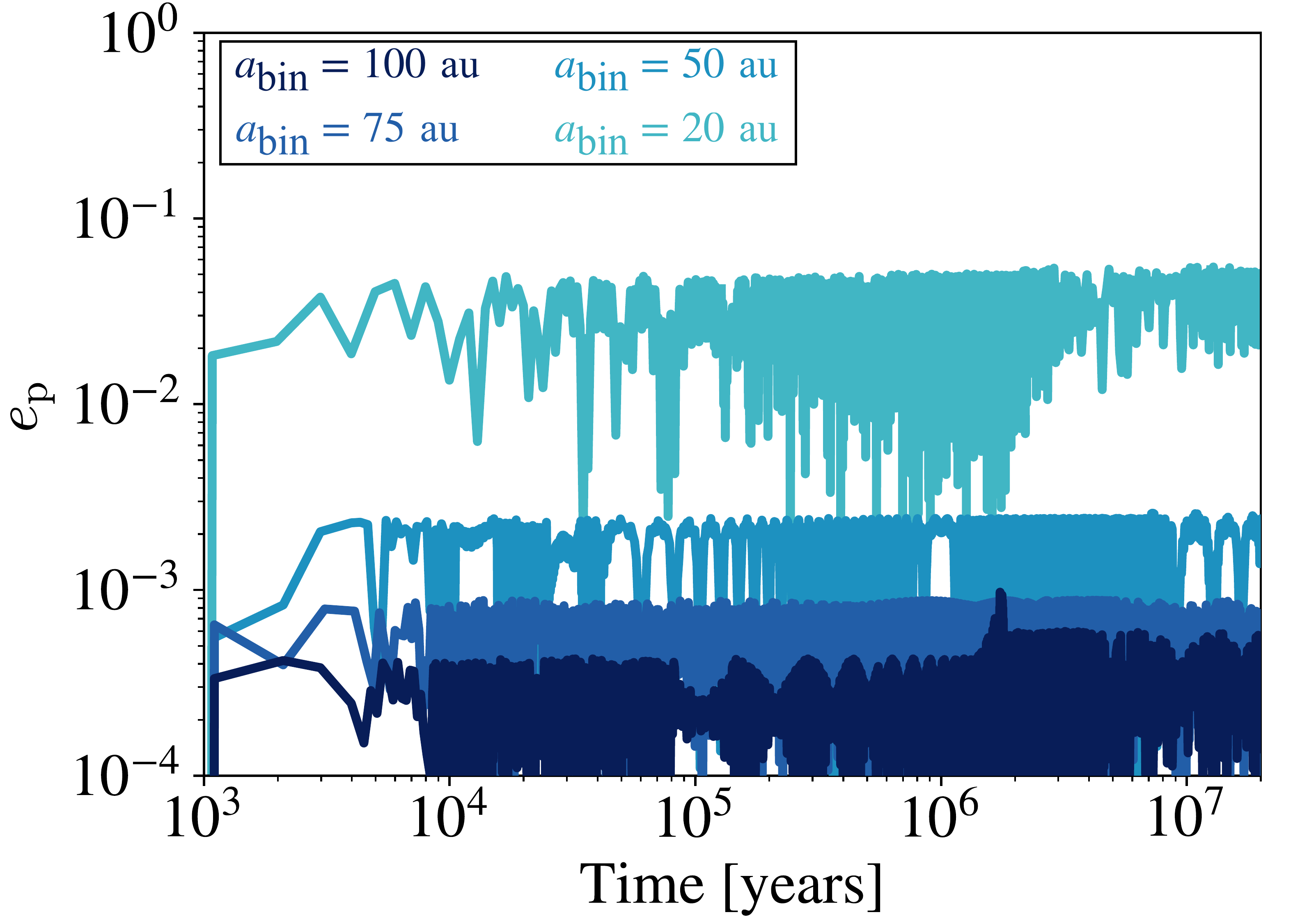}
      
      {\small (c) Eccentricity evolution: embryo at 5 au}
    \end{minipage}
    \hfill
    \begin{minipage}[t]{0.495\linewidth}
      \centering
      \includegraphics[width=\linewidth]{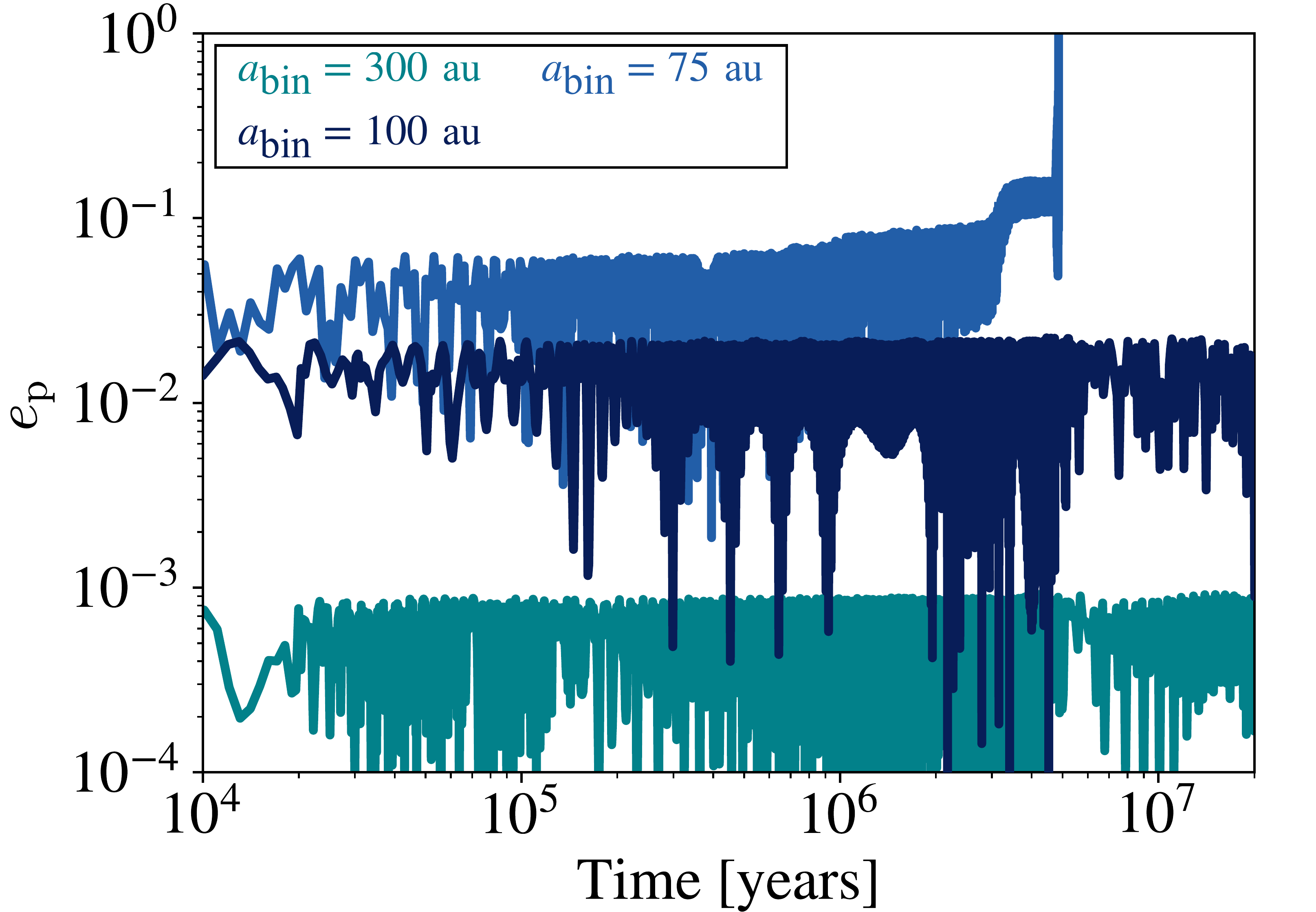}
      
      {\small (d) Eccentricity evolution: embryo at 20 au}
    \end{minipage}
  \end{minipage}
  \hfill
  \begin{minipage}[c]{0.24\textwidth}
    \caption{Simulation set B (in situ): time evolution of planet mass (top row) and eccentricity (bottom row), color-coded by binary separation. The embryo is placed at 5 au (left column) and 20 au (right column). In the top row, the dashed line represents the core mass, while the solid line shows the total planet mass. Note:\ the x-axis and y-axis scales are different in all panels in the top row.}
    \label{insitu_planet_mass_eccentricity}
  \end{minipage}
\end{figure*}

\subsection{Simulation set B: Single versus binary star with the one-embryo scenario}\label{subsec:results_simulationsB}
In simulation set B, we model the growth of a Moon-mass embryo in five different binary configurations (\abin\ = 20, 50, 75, 100, and 300 au; see the middle column of Table \ref{tab:initialconditionsABC} for details). For comparison, we also run a single-star case with identical initial conditions for the central star and disk. In Table \ref{tab:resultsB} we report, for each simulated configuration, the values for binary separation, \abin, truncation radius, $R_\text{trunc}$, critical semimajor axis, $a_\text{crit}$, initial embryo location, $a_\text{p,0}$, and metrics, $\mathcal{M}_{1,2}$. Furthermore, each configuration is tested under two growth scenarios: in situ growth and growth with migration.

\paragraph{In situ}

Figure \ref{insitu_planet_mass_eccentricity} shows the evolution of planet mass (top row) and planet eccentricity (bottom row) over time, color-coded by binary separation. The solid lines in the top row panels represent the total planet mass, while dashed lines indicate the core mass. The embryo is initially placed at 5 au (left column), and 20 au (right column). Notably, in the top left panel, the planet in the \abin\ = 100 au system grows comparably to the single-star case. This occurs because the embryo is located within 1/3 of the Hill sphere radius of the primary star. Conversely, the top right panel shows that in the \abin\ = 100 au system, an embryo at 20 au does not experience significant growth, consistent with our findings from set A, given that $\mathcal{M}_1 = 0.65$ and $\mathcal{M}_2 = 0.33$. Similarly, in the \abin\ = 300 au system (top right panel), the embryo at 20 au exhibits nearly identical growth to the single-star case. Here, the planet is well inside the stable region ($\mathcal{M}_{1,2} < 1/3$) and the secondary star's influence is negligible. In contrast, as the binary separation decreases, the embryos accrete less material overall due to a more compact and less massive protoplanetary disk. In the \abin\ = 75 au case (top right panel), the 20 au embryo fails to grow and is ejected, as indicated by its eccentricity reaching unity (bottom right). This aligns with our findings from set A: as $\mathcal{M}_{2}$ approaches the value of 1/2, planets are more likely to be lost. Furthermore, compared to Paper I, we find that the embryo at 20 au in the \abin\ = 100 au case grows less efficiently when including the gravitational interaction with the secondary star. This is also in line with the results of set A given that this case falls in the region $0.3<\mathcal{M}_{1}<0.4$, where we find a sharp decrease in final planet mass compared to $\mathcal{M}_{1}<0.3$.

\paragraph{With migration}

\begin{figure*}[htb]
  \centering
  \begin{minipage}[t]{0.74\textwidth}
    \centering
    \begin{minipage}[t]{0.48\linewidth}
      \centering
      \includegraphics[width=\linewidth]{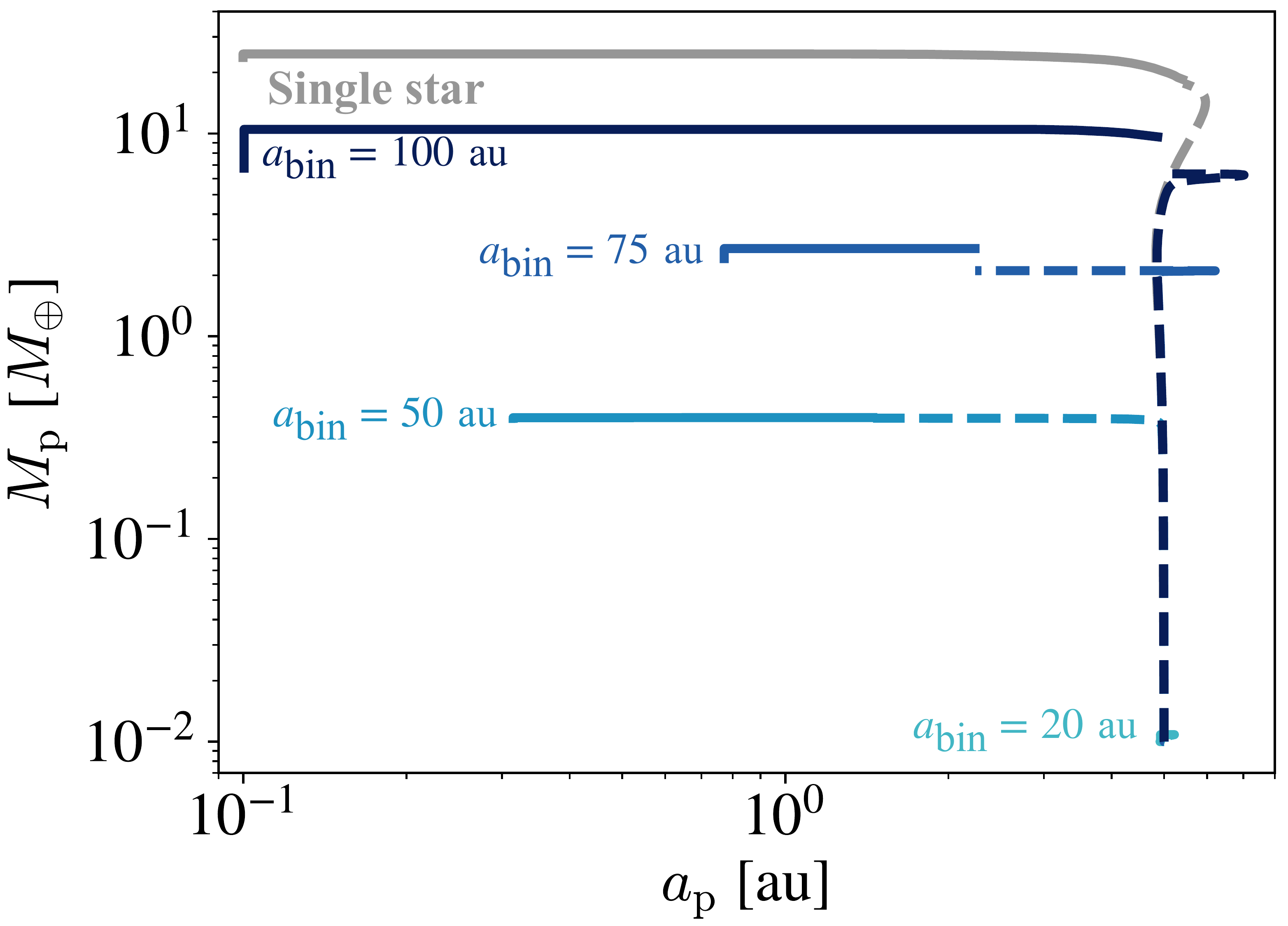}
      
      {\small (a)  Planet formation track: embryo at 5 au}
    \end{minipage}
    \hfill
    \begin{minipage}[t]{0.48\linewidth}
      \centering
      \includegraphics[width=\linewidth]{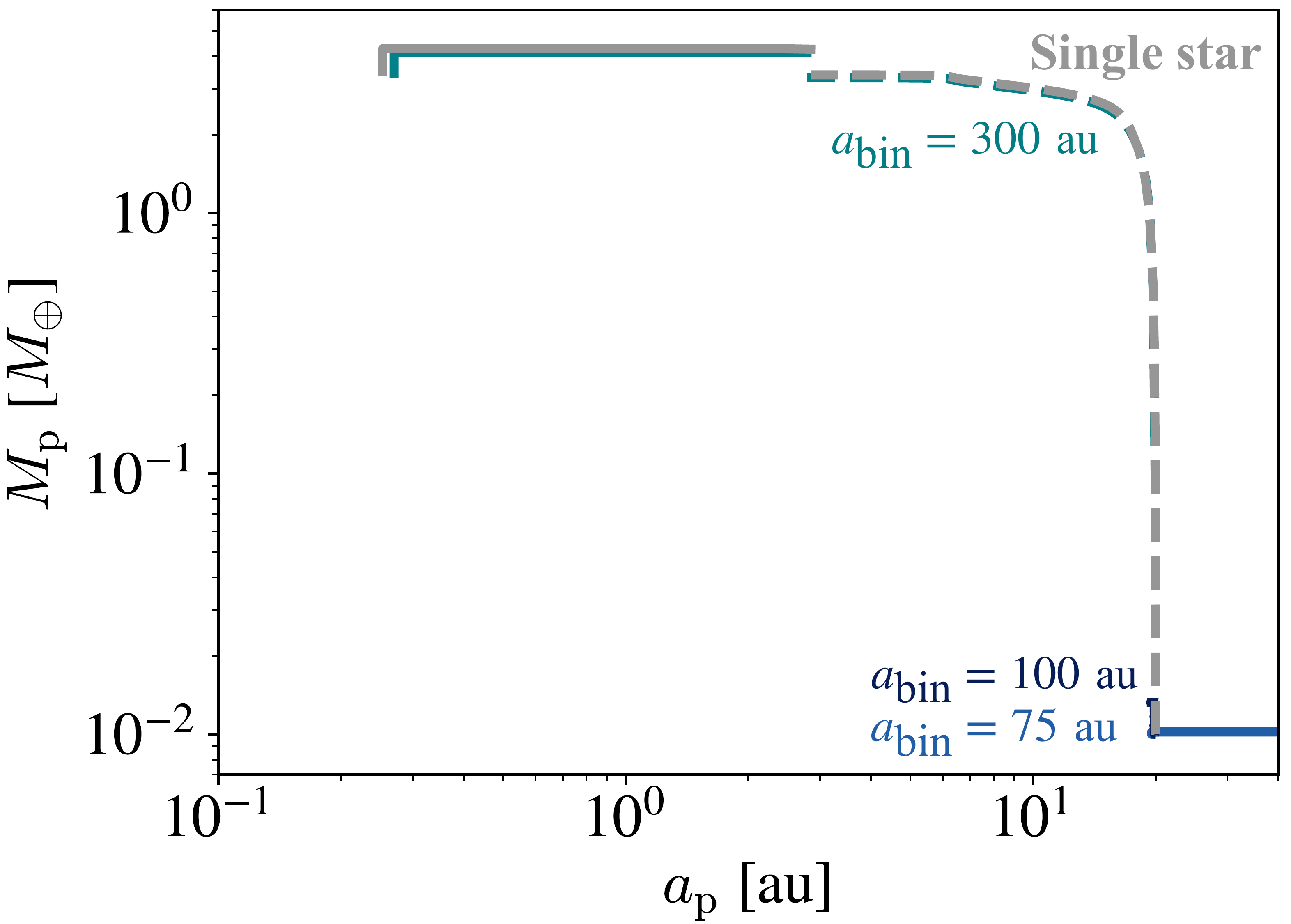}
      
      {\small (b)  Planet formation track:\ embryo at 20 au}
    \end{minipage}
    \\[3mm]
    \begin{minipage}[t]{0.48\linewidth}
      \centering
      \includegraphics[width=\linewidth]{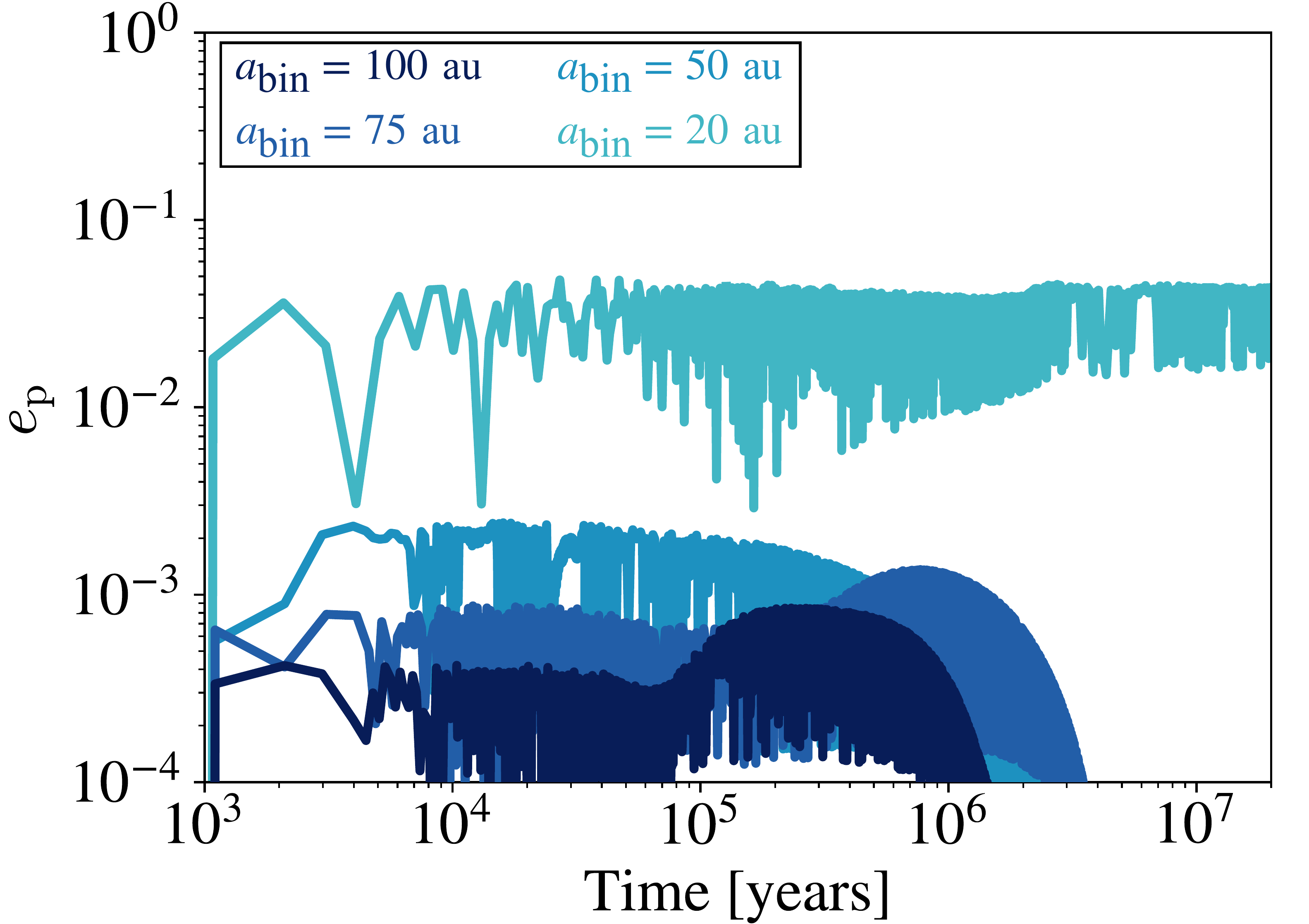}
      
      {\small (c) Eccentricity evolution:\ embryo at 5 au}
    \end{minipage}
    \hfill
    \begin{minipage}[t]{0.48\linewidth}
      \centering
      \includegraphics[width=\linewidth]{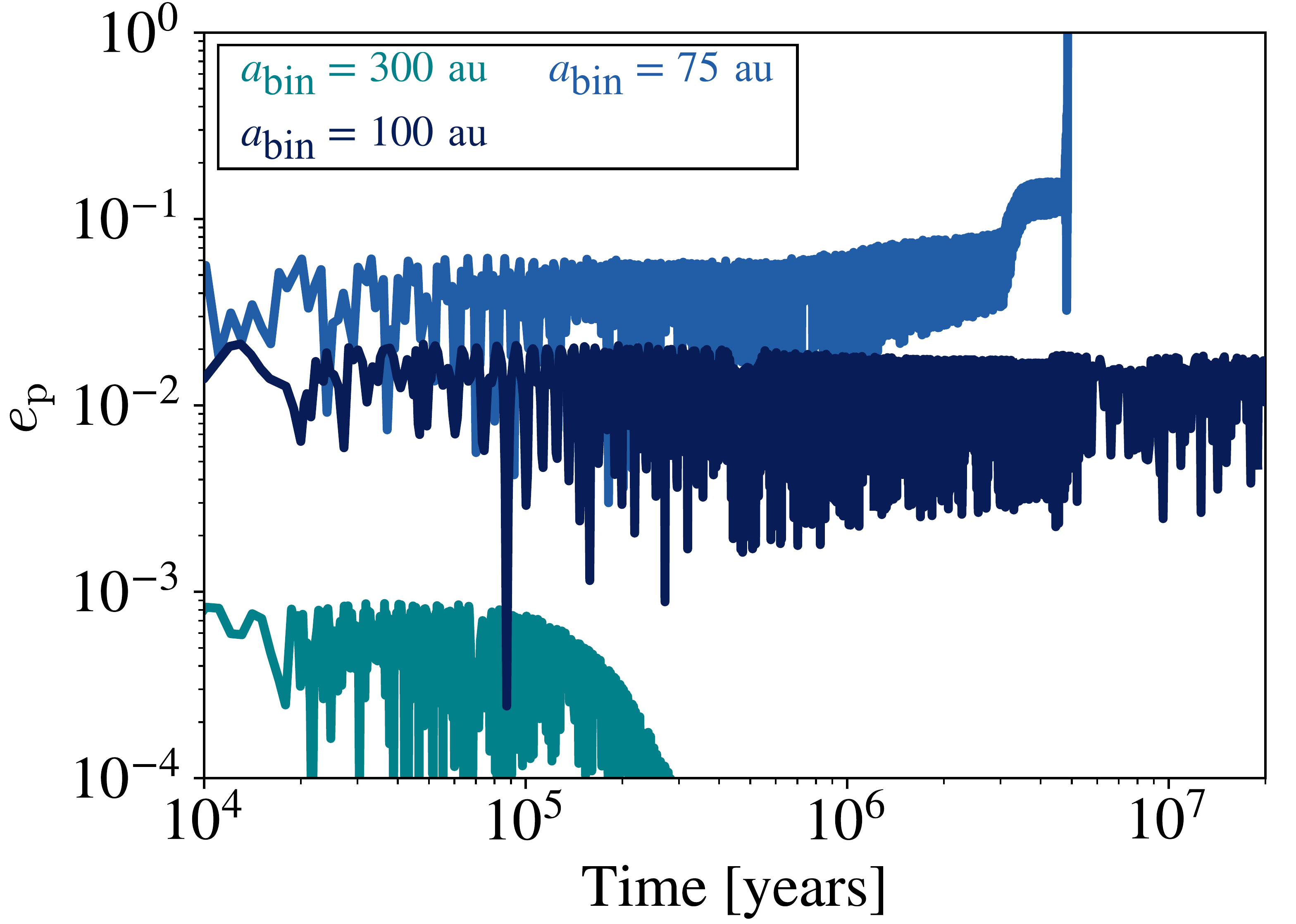}
      
      {\small (d) Eccentricity evolution:\ embryo at 20 au}
    \end{minipage}
  \end{minipage}
  \hfill
  \begin{minipage}[c]{0.24\textwidth}
    \caption{Simulation set B (with migration): semimajor axis versus planet mass (top row) and time versus eccentricity (bottom row), color-coded by binary separation. The embryo is initially placed at 5 au (left column), and 20 au (right column). In the top row, the dashed line represents the core mass, while the solid line shows the total planet mass. Note: the x-axis and y-axis scales are different in all the panels in the top row.}
    \label{migration_SMAvsM_eccentricity}
  \end{minipage}
\end{figure*}

\begin{figure*}[b]
        \centering
        \includegraphics[width=0.67\columnwidth]{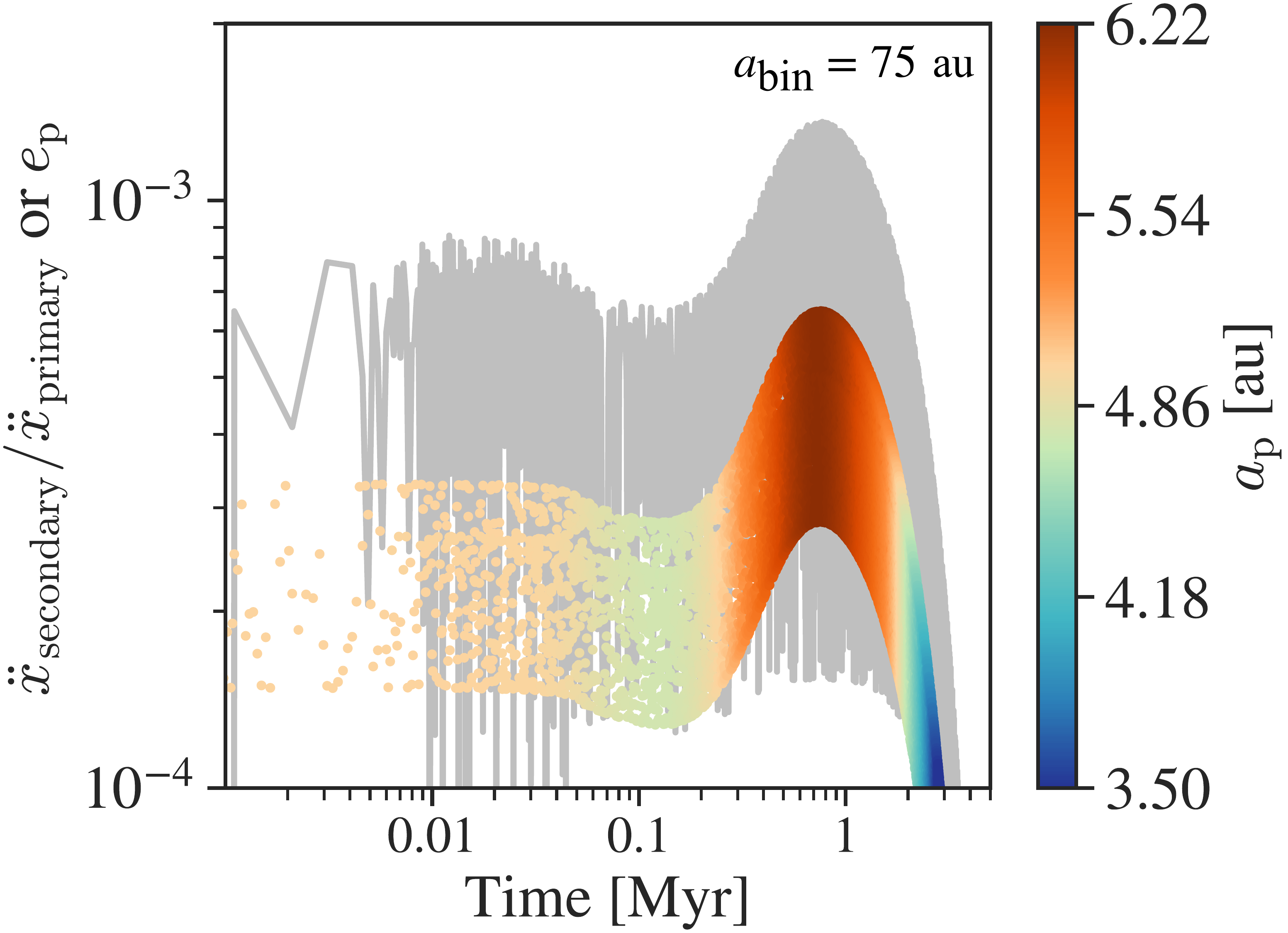}
        \includegraphics[width=0.67\columnwidth]{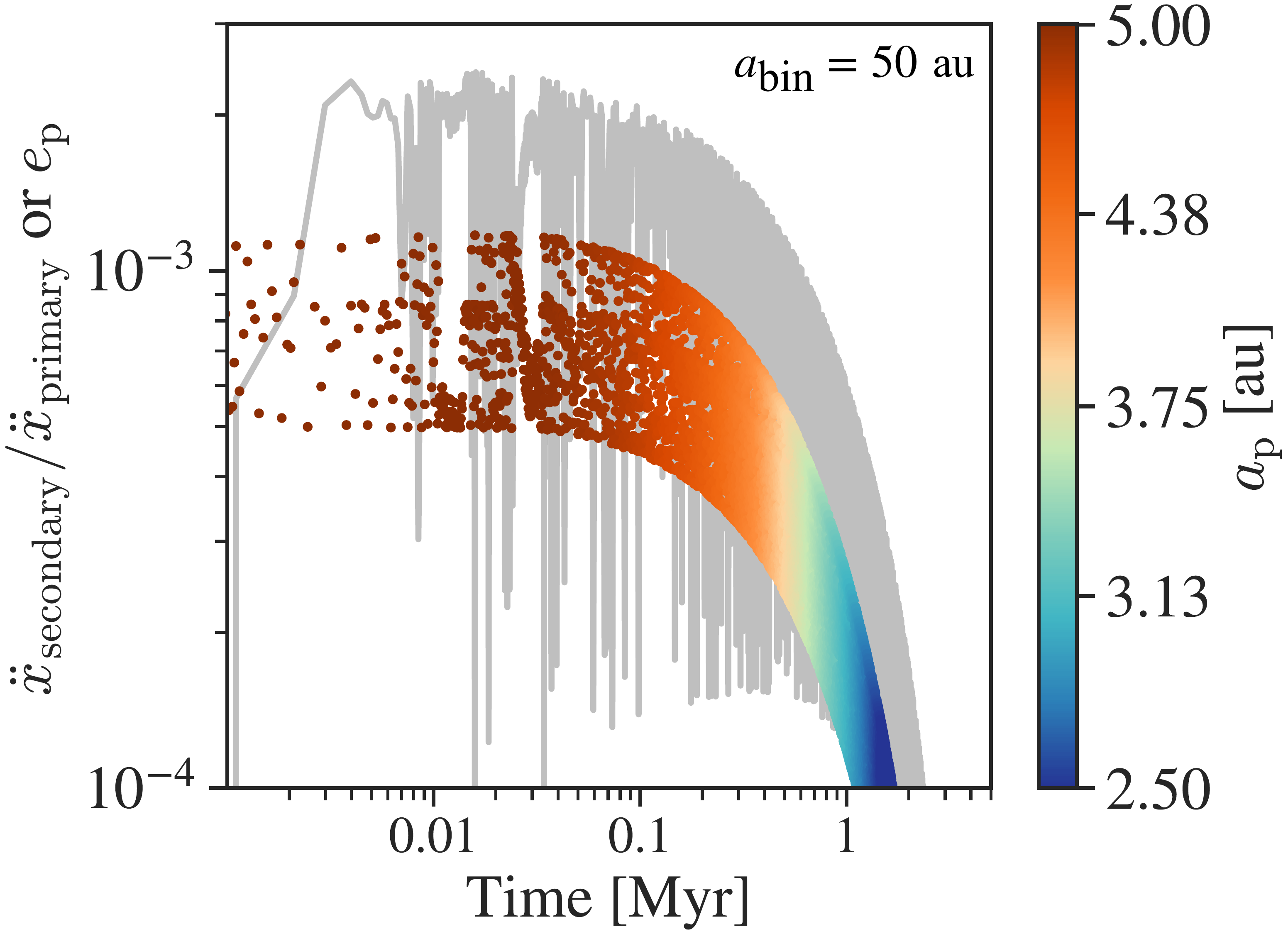}
        \includegraphics[width=0.67\columnwidth]{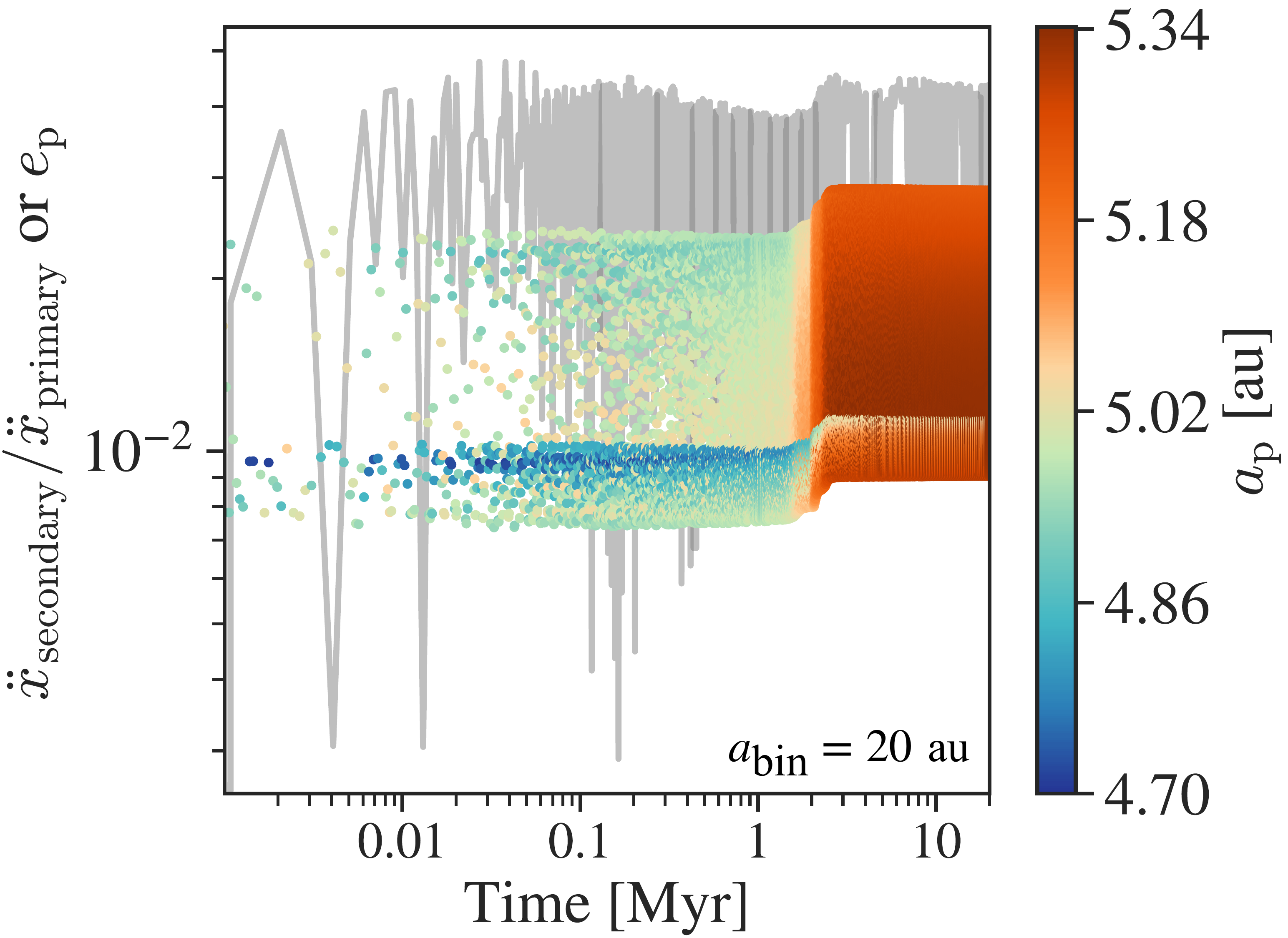}
        \caption{Simulation set B (with migration): time evolution of the planet’s acceleration due to the secondary star ($\ddot{x}_{\text{secondary}}$) normalized by the acceleration due to the primary star ($\ddot{x}_{\text{primary}}$), color-coded by planet semimajor axis. The gray line represents the planet’s eccentricity. Each panel corresponds to a specific binary case: \abin\ = 75 au (left panel), \abin\ = 50 au (middle panel), and \abin\ = 20 au (right panel), with the embryo initially placed at 5 au. Note the different x-axis, y-axis and color bar scales for each plot.}
        \label{acceleration_terms}
\end{figure*}

Figure \ref{migration_SMAvsM_eccentricity} presents the planet's mass as a function of its semimajor axis (top row) and its eccentricity over time (bottom row). The final planet mass and orbital location vary significantly across configurations. In the \abin\ = 20 au (for an embryo at 5 au, top left panel), and \abin\ = 100 au (for an embryo at 20 au, top right panel), migration is minimal compared to other cases. Meanwhile, in the \abin\ = 75 au system (top right panel), the planet starting at 20 au is ejected, as in the in situ scenario. On the other hand, the migration of the planet starting at 5 au in the \abin\ = 75 au system is halted due to the gap formed in the gaseous disk from internal photoevaporation. To further investigate the decrease in the planet eccentricity visible in the middle and right panel of the bottom row of Figure \ref{migration_SMAvsM_eccentricity}, we consider the scenarios in which the planet is initially located at 5 au and we show in Figure \ref{acceleration_terms} the planet's acceleration $\ddot{x}$ due to the secondary star, normalized by the primary star’s acceleration, as a function of time (color-coded by the planet semimajor axis). The gray line represents the planet's eccentricity. We find that in the \abin\ = 75 and 50 au cases (left and middle panels), inward migration reduces the planet's eccentricity by decreasing the gravitational influence of the secondary star. In contrast, in the \abin\ = 20 au system, the planet remains near its initial location and both the secondary star’s perturbation and the planet's eccentricity stay constant (right panel). This behavior is again linked to the embryo's initial placement. When the embryo starts at 5 au in a binary with \abin\ = 20 au, then $\mathcal{M}_{1} = 0.65$ and $\mathcal{M}_{2} = 0.41$. As a result, the planet is unable to accrete enough material to trigger migration, since it resides in the regions where the secondary star’s influence is stronger.

\subsection{Simulation set C: Single versus binary star in the multi-embryo scenario}\label{subsec:results_simulationsC}

To explore how multiple embryos evolve in different environments, we simulated a circular binary with \abin\ = 100 au and a single-star case, each with initially five embryos, placed at the same locations. This allows us to compare the final planetary architectures under identical initial conditions (see the right column of Table \ref{tab:initialconditionsABC} for details).

\begin{figure}[h!]
        \centering
        \includegraphics[width=0.8\columnwidth]{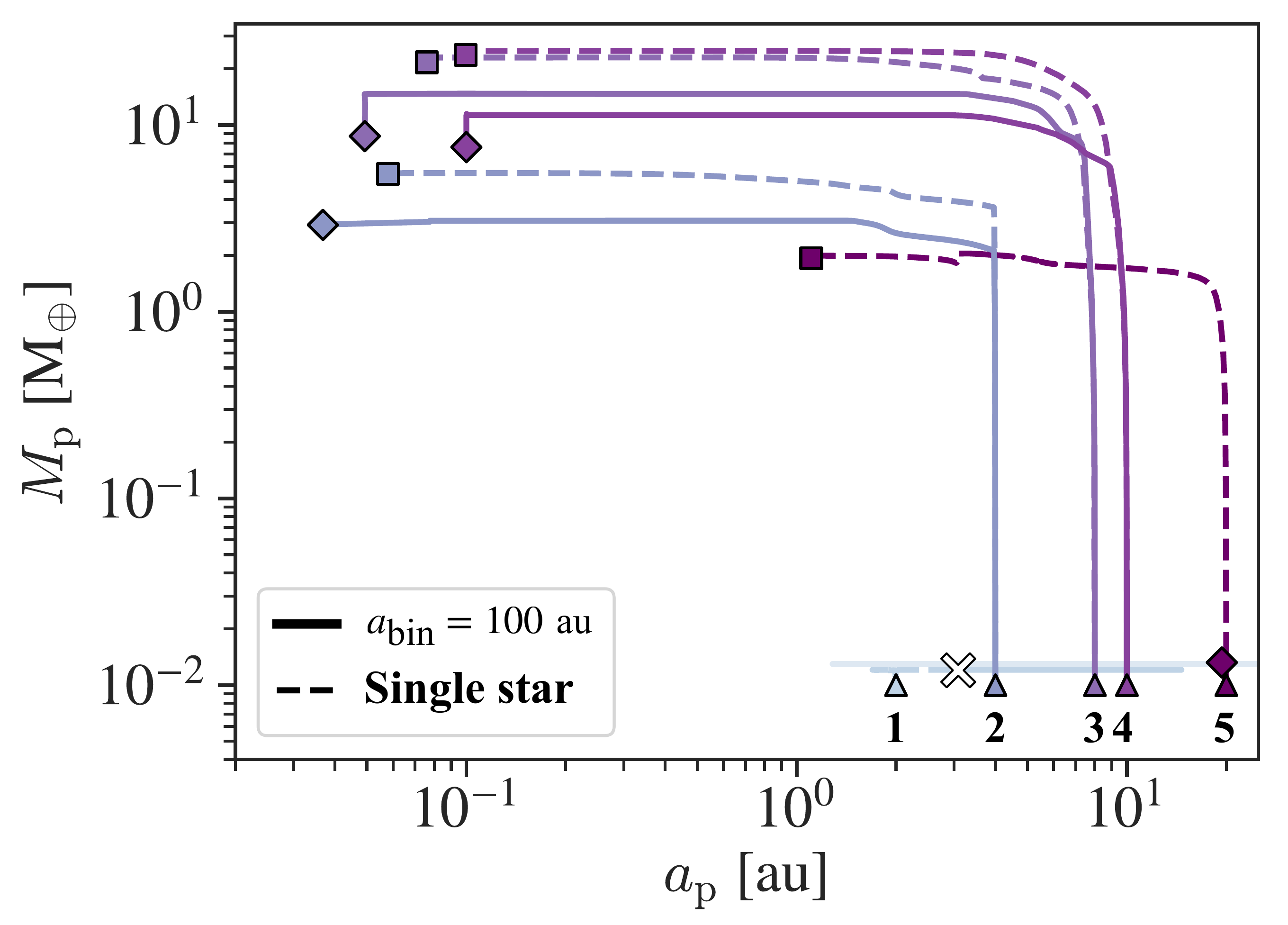}
        \vspace{-0.3cm}
        \caption{Simulation set C: Semimajor axis versus planet mass. The solid lines corresponds to \abin\ = 100 au, while the dashed lines show the single-star scenario. The cross marker indicates a planet-planet collisions between planet 1 and planet 5 from the single-star scenario. In the binary scenario planet 1 is ejected. Triangle markers show the initial planet locations along with their ID numbers, while diamond and square markers represent the final planet masses and semimajor axes in the binary and single-star scenario, respectively.}
        \label{multi_embryo_SMAvsM_single_vs_100au}
\end{figure}

Figure \ref{multi_embryo_SMAvsM_single_vs_100au} shows the planet tracks (semimajor axis vs. mass) for both cases. Solid lines denote the binary case, while dashed lines represent the single-star case. The initial locations of the embryos are marked by triangles and the final masses and semimajor axes are denoted by diamonds and squares for the binary and the single-star scenario, respectively. Each color represents a different embryo. The cross marker indicates the merging between planet \#1 and planet \#5, which looses its envelope after the collision (small dip in the track of planet \#5 at $\sim 3$ au in the single-star scenario). Previously, we found that the single-star and \abin\ = 100 au cases produce similar results when the embryo is located far from the disk truncation radius ($\mathcal{M}_{1}<0.4$, $\mathcal{M}_{2}<1/3$). However, introducing multiple embryos reveals key differences between the two scenarios. In the binary case, the outermost embryo fails to grow, consistent with the single-embryo result at 20 au (Fig. \ref{migration_SMAvsM_eccentricity}, top right panel). Additionally, the innermost embryo is ejected from the system, whereas in the single-star case, it merges with another embryo. By the end of the simulations, both the single-star system and the binary system retain four of their initial bodies. However, in the binary scenario, only three of them are able to reach masses above 0.1 \Mearth. Furthermore, the more massive planets in both systems end up at different respective orbital distances from their star. This highlights the impact of binary interactions on the resulting S-type planetary architecture.

\section{Discussion}\label{sec:discussion}

Our results highlight the importance of considering the gravity of the secondary star when modeling the formation of S-type planets. In simulation set A, we performed in situ one-embryo planet formation simulation, exploring a wide range of binary architectures and initial embryo locations. We find that planetary survival and growth are strongly linked to the relative position of the embryo with respect to both the disk truncation radius and the primary star’s Hill radius. Planets placed beyond $\mathcal{M}_1 \sim 0.4$ (i.e., outside 40\% of the truncation radius) are increasingly subject to dynamical instabilities and their loss can be directly understood in terms of their distance from the Hill radius of the primary star. Planets placed beyond one-third of the primary star’s Hill radius ($\mathcal{M}_2 > 1/3$) start to experience instabilities and, when the planet is placed beyond half of the primary star's Hill radius ($\mathcal{M}_2 > 1/2$), planets are more likely to be lost (Figure \ref{insitu_grid1}). Quantitatively, 28.6\% of planets with $1/3 < \mathcal{M}_2 < 1/2$ and 86.5\% of planets with $\mathcal{M}_2 > 1/2$ become unbound. This trend is in line with the findings of \cite{Domingos2006}, who studied the stability of extrasolar planets' satellites, which is another example of a hierarchical three-body problem. These authors found that the critical semimajor axis, beyond which satellite orbits are not stable is approximately half of the Hill radius of the orbited body. In terms of planet stability, our results are also consistent with the stability criteria proposed by \cite{Quarles2020}, as we find that planets with $a_{\text{p,0}} > a_\text{crit}$ are more likely to become unstable. Quantitatively, 78.4\% of planets located beyond this limit are ultimately lost. While there is no clear trend with the binary mass ratio alone, binary eccentricity plays an equally central role in shaping planetary orbits. Systems with high $e_\text{bin}$ tend to excite larger planetary eccentricities (Figure \ref{insitu_grid2}), consistent with the analytic expression for equilibrium eccentricity derived by \citet{Mardling2007}. In fact, 58.9\% of the lost planets were hosted by binaries with $e_\text{bin}>0.5$. When planets do survive beyond $\mathcal{M}_1 \sim 0.4$, their ability to grow is nevertheless severely compromised. In the inner parts of the disk ($\mathcal{M}_1 \lesssim$ 0.4, but especially $\mathcal{M}_1 \lesssim$ 0.3), planets can still reach several Jupiter masses, provided that the binary periastron distance is wider than 20 au (Figure \ref{insitu_grid3}). However, as the embryo is placed further out in the disk, the combined effect of disk truncation and stellar gravitational perturbations sharply reduces the growth efficiency. Compared to the results in Paper I, which only accounted  for disk truncation, we find a dramatic reduction in the mean planet mass already at $0.3<\mathcal{M}_1<0.4$ (26 \Mearth\ vs. 44 \Mearth, a decrease of about 41\%). 
Beyond $\mathcal{M}_1 \gtrsim 0.4$, the reduction is even more pronounced: while the study in Paper I still produced 136 planets more massive than 0.1 \Mearth, in the present work, all planets except two remain at their initial mass of $\sim 0.01$ \Mearth. This highlights that disk truncation alone does not fully account for the reduced efficiency of planet growth in binaries. It is essential to also consider the gravitational perturbations from the secondary star, which not only affect the planet’s stability but also hinder its ability to accrete material by increasing its orbital eccentricity. Eccentricity influences the pebble accretion rate through the relative velocity between the planet and the pebbles, as described in Eq. (36) of \cite{johansenandlambrechts17}. As a result, pebble accretion becomes significantly less efficient for eccentricities around or above 0.1 (see also top panel of Fig. \ref{insitu_grid1}). 
 
In set B, we analyzed six specific binary cases and compared them to a single-star scenario with identical initial conditions for the protoplanetary disk and embryo location. The results align with the trends observed in set A. We find that planets forming with $\mathcal{M}_2 < 1/3$ can grow and if solid accretion and migration are not significantly hindered by the reduced material supply due to disk truncation, some planets can reach masses and final locations similar to those in single-star systems, just as in the case of our Paper I (5 au embryo in the \abin\ = 100 au case and 20 au embryo in the \abin\ = 300 au case). In contrast, planets with $\mathcal{M}_2 > 1/3$ experience strong perturbations from the secondary star, limiting their accretion and migration and often leading to instabilities, particularly when $\mathcal{M}_2 > 1/2$. Notably, the embryo at 20 au in the \abin\ = 100 au case grows less efficiently when it experiences the gravitational interaction with the secondary star and the embryo at 20 au in the \abin\ = 75 au case is ejected from the system, as compared to the case of Paper I (panels (b) and (d) of our Figure \ref{insitu_planet_mass_eccentricity}). 

Lastly, in simulation set C, we compare one illustrative binary case with a single star case, revealing that even if the embryos are placed within 1/3 of the primary star's Hill radius, in the context of multi-embryo simulations, the outcomes between S-type binary and single star case are not identical due to the chaotic nature of the process (Figure \ref{multi_embryo_SMAvsM_single_vs_100au}). While in the case of the single-star system, both systems ultimately retain four bodies, in the binary case, one of them is not able to accrete any solid material. Furthermore, the final planet masses and orbital locations differ. This highlights the interplay between binary perturbations and planet-planet interactions. 

\subsection{Dependence on the disk parameters}
\begin{figure}[h!]
        \centering
        \includegraphics[width=0.9\hsize]{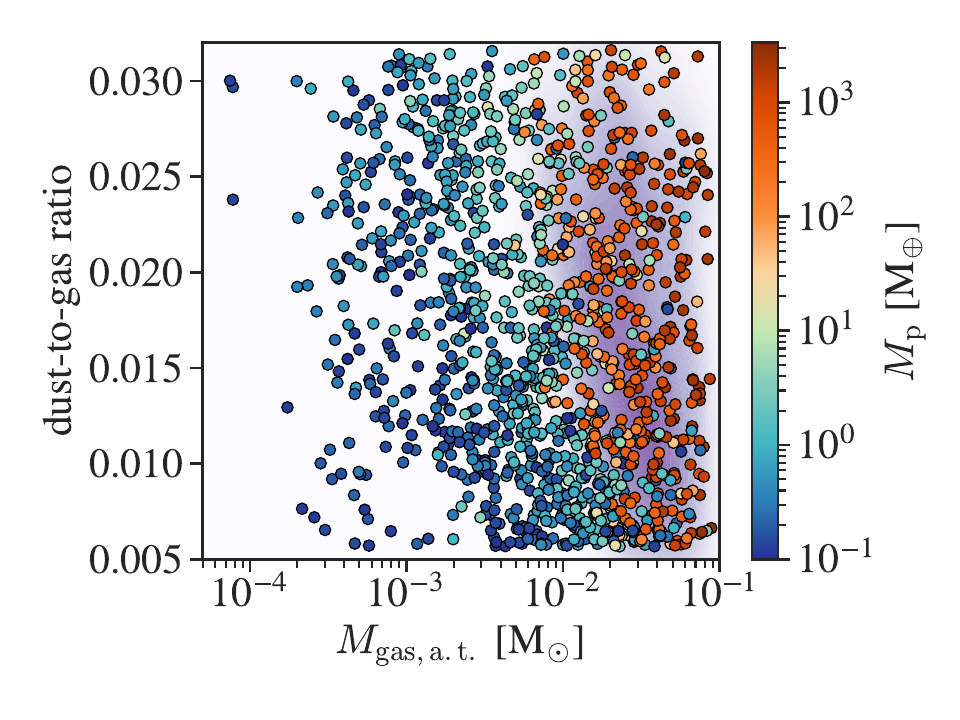}
        \includegraphics[width=0.9\hsize]{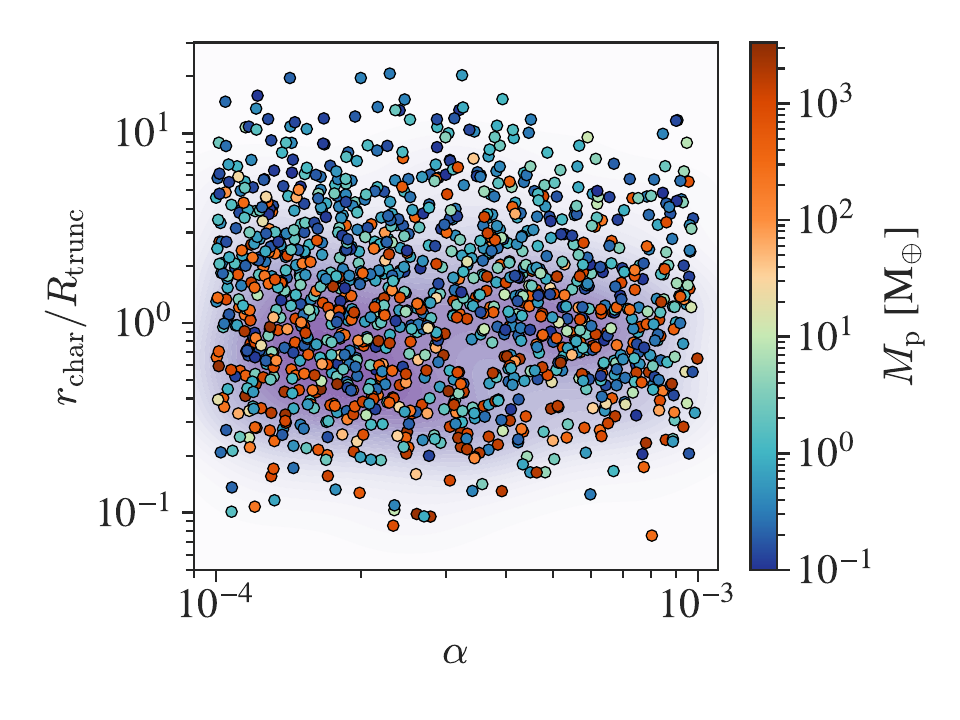}
        \vspace{-0.5cm}
        \caption{Simulation set A: Disk mass after truncation versus dust-to-gas ratio (top panel) and disk viscosity parameter versus the ratio between the disk characteristic radius and the disk truncation radius, $r_{\rm char}/R_{\rm trunc}$ (bottom panel). We show planets that became at least more massive than Mars and their color reflects their final planet mass. The shaded contours highlight regions of highest planet density, obtained through a two-dimensional kernel density estimation (KDE) considering planets with masses above 10 \Mearth.}
        \label{disc_dependance}
\end{figure}

In simulation set A, in addition to sampling the binary parameters, we also sampled the disk properties ($M_{\rm gas, b.t.}$, dust-to-gas ratio, $\alpha$, $r_{\rm char}$). We find that the disk parameters do not affect the dynamical stability of the planets, but they do affect their final masses. The top panel of Figure \ref{disc_dependance} shows the dust-to-gas ratio as a function of the disk gas mass after truncation ($M_{\rm gas, a.t.}$), while the bottom panel displays the ratio between the characteristic disk radius and the truncation radius ($r_{\rm char}/R_{\rm trunc}$) as a function of the disk viscosity parameter, $\alpha$. In both panels, only planets that reached at least the mass of Mars are plotted, and the color scale represents their final mass. The purple contour highlights the region with the highest density of planets more massive than 10 \Mearth. We plot $r_{\rm char}/R_{\rm trunc}$ instead of $r_{\rm char}$ alone to test whether, for a given $M_{\rm gas, b.t.}$ and $R_{\rm trunc}$, disks with $r_{\rm char} < R_{\rm trunc}$ lead to different final planet masses compared to those with $r_{\rm char} > R_{\rm trunc}$. However, no significant trend is found in this respect, nor with respect to the dust-to-gas ratio or the viscosity parameter $\alpha$. The parameter with the strongest influence is the disk gas mass: values above 0.01 \Msun\ after truncation are generally required to form most of the planets with $M_{\rm p}$ > 10 \Mearth.

\subsection{Model limitations}
A first limitation of the model concerns the treatment of collisions between planetary embryos. Collisions are modeled as perfect mergers, in which the cores are fully accreted and the gaseous envelopes are entirely ejected. This simplified prescription neglects the possibility of erosive events, which could affect the final planetary masses and compositions. Another limitation arises from the temporal extent of the N-body integration, which is performed up to 20 Myr. The choice of not extending the N-body calculations for Gyrs is due to high computational cost, especially when multiple embryos are included. Finally, although the N-body integrator can handle inclined binary configurations, only coplanar systems are considered in this study. Introducing an inclination would require a more complex treatment of the disk’s structure, particularly regarding the irradiation induced by the stellar companion. All these limitations are beyond the scope of the present work, but could be addressed in future developments.

\section{Conclusions}\label{sec:conclusions}

In this work, we introduce, for the first time, the effect of the secondary star’s gravity in a global planet formation model for S-type planets. While Paper I focused on the effect of the disk modifications alone, here, we explicitly account for the gravitational perturbations exerted by the stellar companion on planetary embryos. Our results demonstrate that incorporating these binary-specific interactions is essential for accurately modeling the unique environment where S-type planets form.

We find that planets placed beyond 40\% of the disk truncation radius can become unstable without being able to accrete any material from the disk (Figure \ref{insitu_grid1}). This threshold is related to the planet’s position relative to the primary star’s Hill radius. Indeed, planet eccentricity and the frequency of planet losses increase as the planet approaches and exceeds half the Hill radius, especially when the binary eccentricity is high (\ebin\ > 0.5). Approximately six out of seven planets that are beyond one-half of their host star’s Hill radius are lost from the system. Even when embryos are initially placed within the dynamically stable region, their growth and final architecture are shaped not only by the reduced material supply (due to disk truncation), but also by the dynamical imprint of the binary companion, which excites planetary eccentricities (Figure \ref{insitu_grid2}). This excitation hinders pebble accretion by raising the relative velocity between planets and pebbles. Compared to the case where only disk truncation is considered, the additional effect of stellar perturbations entirely suppresses in situ growth beyond 0.5$\ R_{\rm trunc}$ (Figure \ref{insitu_grid3}). Specifically, all planets more massive than Mars form in situ within 0.5$\ R_{\rm trunc}$, while most planets with $M_{\rm p}$ > 10 \Mearth\ form when $a_{\rm p}<0.3\ R_{\rm trunc}$ (Figure \ref{insitu_grid4}). In terms of \abin\ it means that an equal mass, circular binary with \abin=100 au would form all planets more massive than Mars within $\sim$15 au, with most of planets more massive than Earth within $\sim$10 au. Moreover, the mass of gas available in the truncated disk plays a role in enabling the formation of massive planets: values above $\sim$0.01 \Msun\ are generally required to form the majority of planets more massive than 10 \Mearth. Lastly, even within stable regions and with identical initial conditions, planet formation outcomes differ from those in single-star systems when multiple embryos are included, due to the combined effects of binary perturbations and planet-planet interactions (Figure \ref{multi_embryo_SMAvsM_single_vs_100au}).

Altogether, our findings suggest that modeling planet formation in binary systems requires a comprehensive treatment that includes both the modified disk structure and the gravitational effects of the stellar companion. Ignoring either of these aspects can lead to a substantial mischaracterization of the resulting planetary systems and of the conditions that have shaped the formation of real S-type exoplanets. To further confirm these findings and thoroughly address planet formation in S-type binaries, it will be crucial to perform a population synthesis and perform a statistical comparison with the observed S-type planet population.

\begin{acknowledgements}
    We thank C. Mordasini, A. Kessler, J. Weder, N. Kaufmann and R. Burn for valuable discussions about the Bern Model. We also thank the anonymous referee for providing useful feedback that enhanced the completeness of our work.
    A.N. and J.V. acknowledge support from the Swiss National Science Foundation (SNSF) under grant PZ00P2\_208945.
    D.T. acknowledges support by the Fondazione ICSC, Spoke 3 ``Astrophysics and Cosmos Observations'', National Recovery and Resilience Plan (Piano Nazionale di Ripresa e Resilienza, PNRR) Project ID CN\_00000013 ``Italian Research Center on High-Performance Computing, Big Data and Quantum Computing'' funded by MUR Missione 4 Componente 2 Investimento 1.4: Potenziamento strutture di ricerca e creazione di “campioni nazionali di R\&S (M4C2-19)'' - Next Generation EU (NGEU). D.T. also acknowledges support from the ASI-INAF grant no. 2021-5-HH.0 plus addenda no. 2021-5-HH.1-2022 and 2021-5-HH.2-2024, the COST Action CA22133 PLANETS and the European Research Council via the Horizon 2020 Framework Programme ERC Synergy ``ECOGAL'' Project GA-855130 
    E.B. acknowledges the financial support of the SNSF (grant number: 200021\_197176 and 200020\_215760). This work has been carried out within the framework of the NCCR PlanetS supported by the Swiss National Science Foundation under grants 51NF40\_182901 and 51NF40\_205606. The computations were performed at University of Geneva on the Yggdrasil cluster. This research has made use of the Astrophysics Data System, funded by NASA under Cooperative Agreement 80NSSC25M7105. 
\end{acknowledgements}

\bibliographystyle{aa} 
\bibliography{Biblio}

\begin{appendix}\label{sec:appendix}

\section{S-type N-body integrator}
\subsection{Coordinate transformations}\label{sec:coordinate_trasformations}
In this section we provide the coordinate transformations used in our N-body integrator, which are not explicitly provided in \cite{Chambers99,Chambers02}. For the detailed description of the integrator we refer to Section \ref{sec:symplStype} and the original works of \cite{Chambers99,Chambers02}.  The momenta, $\vec{P,}$ are related to the pseudo-velocities, $\vec{V}$, rather than the real velocities $\dot{\vec{X}}$, and \citet{Verrier} writes their relation as
\begin{equation}\label{P&pseudoV}
    \begin{split}
        &\vec{P}_A =m_{\text{tot}}\vec{V}_A,\\
        &\vec{P}_B =m_{B}\vec{V}_B,\\
        &\vec{P}_i =m_{i}\vec{V}_i.\\
    \end{split}
\end{equation}
\begin{equation}\label{V&pseudoV}
    \begin{split}
        &\vec{V}_A =\dot{\vec{X}}_A,\\
        &\vec{V}_B =\dfrac{m_{Ap}}{m_{\text{tot}}}\dot{\vec{X}}_B,\\
        &\vec{V}_i =\dot{\vec{X}}_i-\dfrac{\sum_j m_j\dot{\vec{X}}_j}{m_{\text{Ap}}},\\
        &\dot{\vec{X}}_i = \vec{V}_i+\dfrac{\sum_j m_j\vec{V}_j}{m_A},
    \end{split}
\end{equation}
The inverse transformation, from S-type coordinates $(\vec{X},\vec{V})$ to inertial coordinates $(\vec{x},\vec{v})$, are also given by \cite{Verrier}:
 \begin{equation}\label{Xtoinertial}
     \begin{split}
         &\vec{x}_A = \vec{X}_A-\dfrac{m_B}{m_{\text{tot}}}\vec{X}_B-\dfrac{\sum_j m_j\vec{X}_j}{m_{\text{Ap}}},\\
         &\vec{x}_i =\vec{X}_A+\vec{X}_i-\dfrac{\sum_j m_j\vec{X}_j}{m_{\text{Ap}}},\\
         &\vec{x}_B = \vec{X}_A-\dfrac{m_{\text{Ap}}}{m_{\text{tot}}}\vec{X}_B.
     \end{split}
 \end{equation}
 \begin{equation}\label{Vtoinertial}
     \begin{split}
         &\vec{v}_A = \vec{V}_A-\dfrac{m_B}{m_{\text{Ap}}}\vec{V}_B-\dfrac{\sum_j m_j\vec{V}_j}{m_A},\\
         &\vec{v}_i =\vec{V}_i+\vec{V}_A-\dfrac{m_B}{m_{\text{Ap}}}\vec{V}_B,\\
         &\vec{v}_B = \vec{V}_A+\vec{V}_B.
     \end{split}
 \end{equation} 
The coordinate transformations from S-type to heliocentric coordinates (i.e., with respect to the primary star) are the following: 
\begin{equation}\label{Xtohelio}
    \begin{split}
        &\vec{x}_{B,helio}=\vec{x}_B-\vec{x}_A=\vec{X}_B+\dfrac{\sum_j m_j\vec{X}_j}{m_{\text{Ap}}},\\
        &\vec{x}_{i,helio}=\vec{x}_i-\vec{x}_A=\vec{X}_i.
    \end{split}
\end{equation}
\begin{equation}\label{Vtohelio}
    \begin{split}
        \vec{v}_{B,helio} &=\vec{v}_B-\vec{v}_A=\dfrac{m_{\text{tot}}}{m_{\text{Ap}}}\vec{V}_B+\dfrac{\sum_j m_j\vec{V}_j}{m_A},\\
        \vec{v}_{i,helio} &=\vec{v}_i-\vec{v}_A=\vec{V}_i+\dfrac{\sum_j m_j\vec{V}_j}{m_A}.
    \end{split}
\end{equation}
The coordinate transformations from heliocentric coordinates to S-type coordinates are
\begin{equation}\label{heliotoX}
    \begin{split}
        \vec{X}_B &=\vec{x}_{B,helio}-\dfrac{\sum_j m_j\vec{x}_{i,helio}}{m_{\text{Ap}}},\\
        \vec{X}_i &= \vec{x}_{i,helio.}
    \end{split}
\end{equation}
\begin{equation}\label{heliotoV}
    \begin{split}
        \vec{V}_B &= \dfrac{m_{\text{Ap}}}{m_{\text{tot}}}\vec{v}_{B,helio}-\dfrac{\sum_j m_j\vec{v}_{j,helio}}{m_{\text{tot}}},\\
        \vec{V}_i &= \vec{v}_{i,helio}-\dfrac{\sum_j m_j\vec{v}_{j,helio}}{m_{\text{Ap}}}.
    \end{split}
\end{equation}
\subsection{Testing} \label{sec:Nbody_testing}
\begin{figure*}[h]
        \centering
        \subfloat
        {\includegraphics[width=0.8\columnwidth]{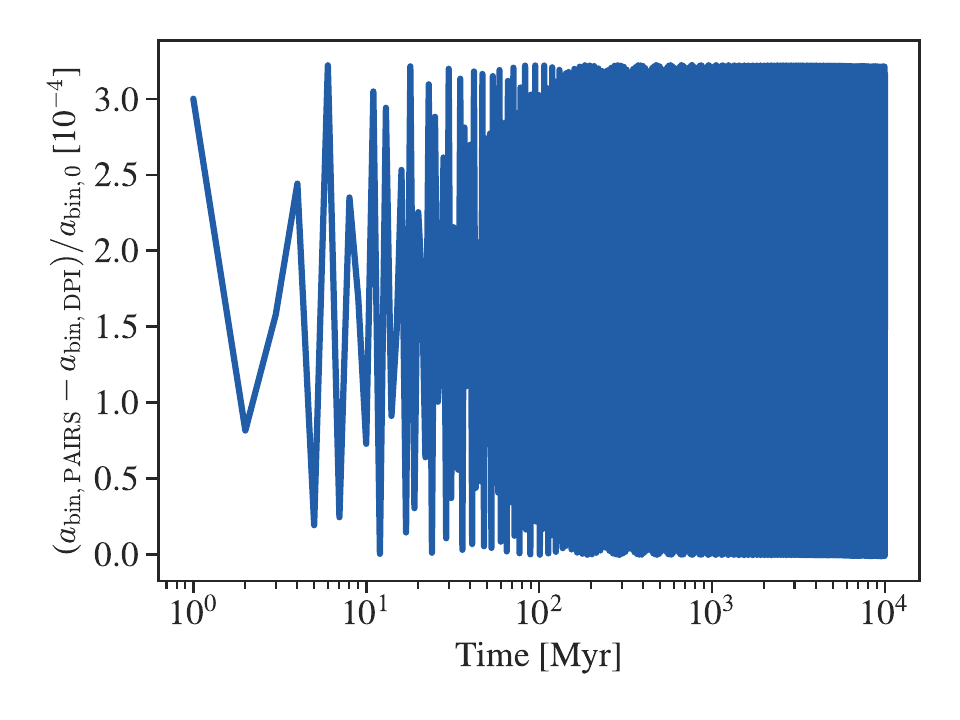}}
        \;\;
        \subfloat
        {\includegraphics[width=0.8\columnwidth]{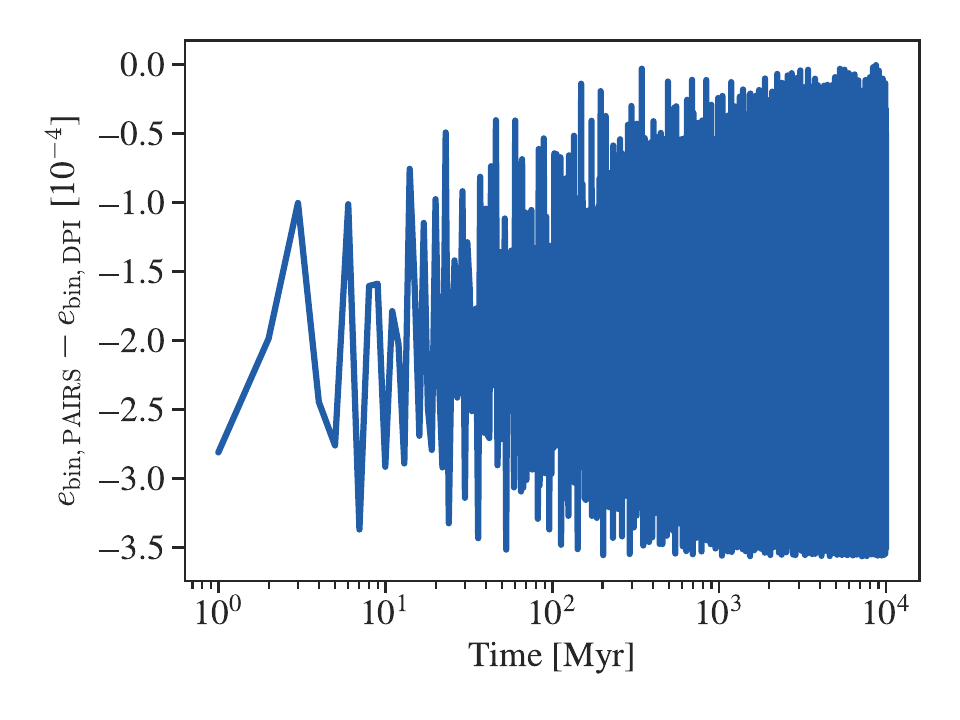}}
        \;\;
        \subfloat
        {\includegraphics[width=0.8\columnwidth]{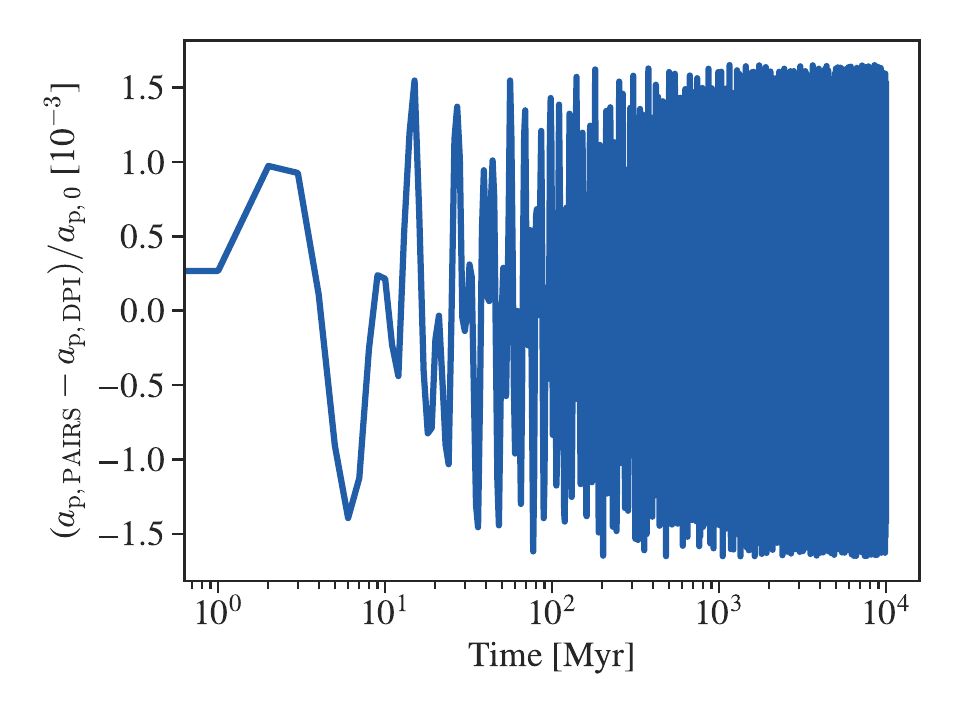}}
        \;\;
        \subfloat
        {\includegraphics[width=0.8\columnwidth]{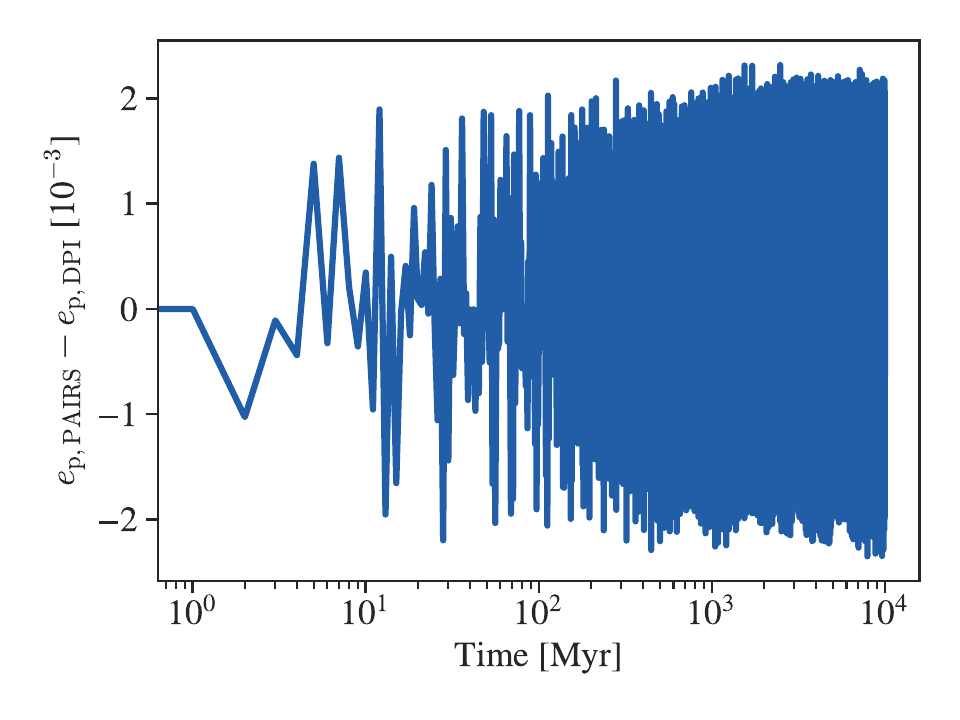}}
        \vspace{-0.5cm}
        \caption{Pure three-body simulations: (top left panel) Difference between the binary semimajor axis calculated with our code and the binary semimajor axis calculated with the DPI library, normalized to the binary semimajor axis of 100 au given as input. (top right panel) Difference between the binary eccentricity calculated with out code and the binary eccentricity calculated with the DPI library. (bottom left panel) Difference between the planet semimajor axis calculated with our code and the planet semimajor axis calculated with the DPI library, normalized to the planet semimajor axis of 10 au given as input. (bottom right panel) Difference between the planet eccentricity calculated with out code and the planet eccentricity calculated with the DPI library.}
    \label{Planete_vs_DPI_1}
\end{figure*}

\begin{figure*}[h]
        \centering
        \subfloat
        {\includegraphics[width=0.8\columnwidth]{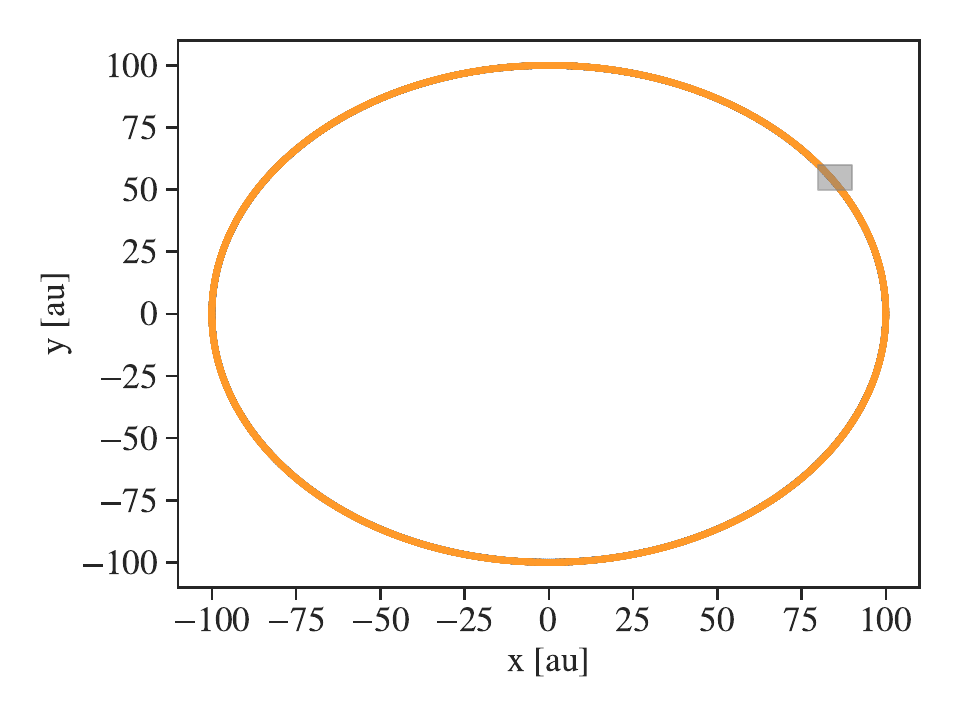}}
        \;\;
        \subfloat
        {\includegraphics[width=0.8\columnwidth]{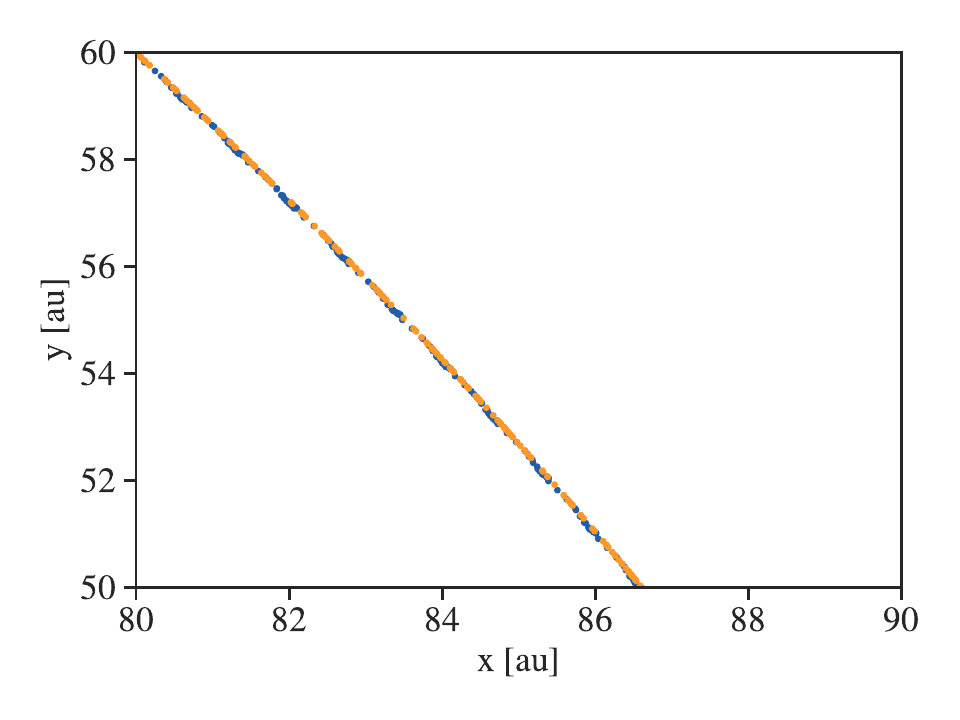}}
        \;\;
        \subfloat
        {\includegraphics[width=0.8\columnwidth]{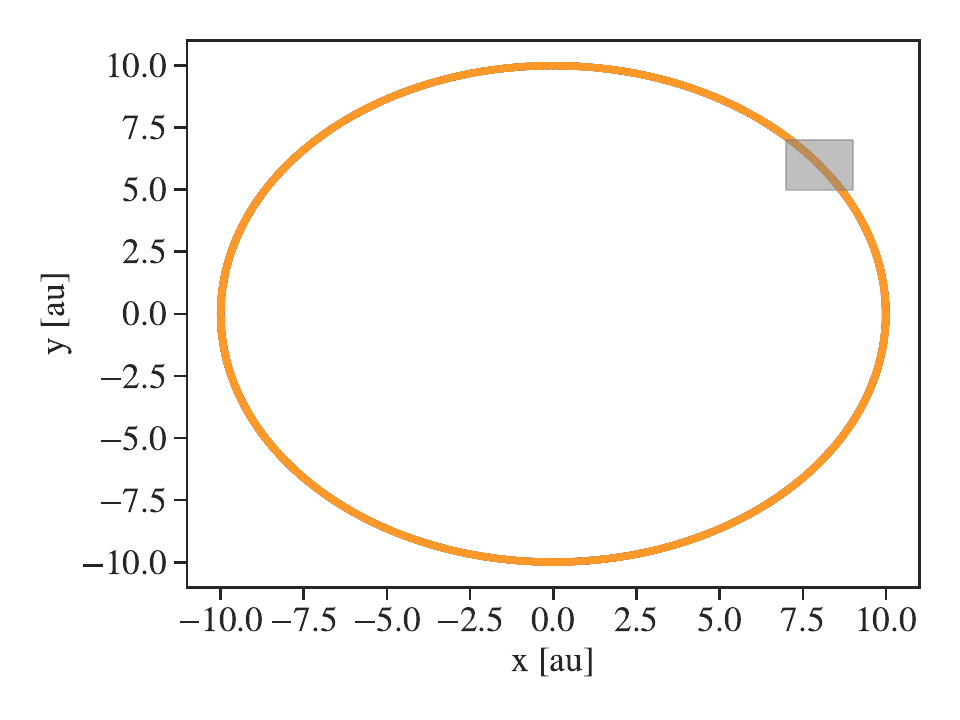}}
        \;\;
        \subfloat
        {\includegraphics[width=0.8\columnwidth]{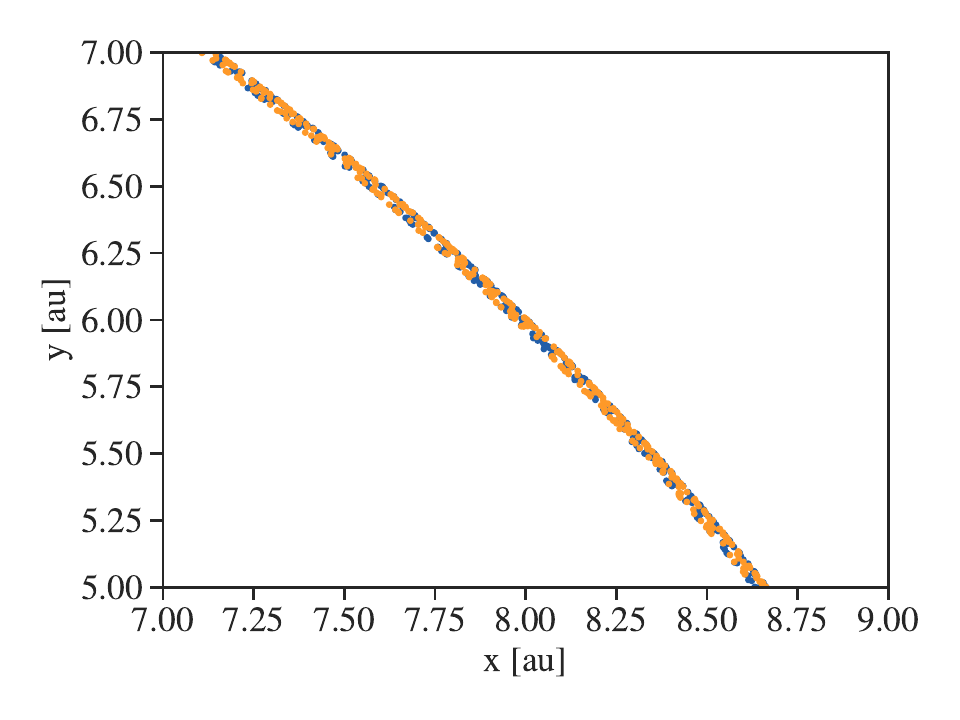}}
        \vspace{-0.5cm}
        \caption{Pure three-body simulations: secondary star's (top panels) and planet's (bottom panels) orbit on the x-y plane. Right:\ Zoomed-in region that is highlighted in gray in the left panels. The simulation run with our code is displayed with an orange color, while the simulation run with the DPI library are in blue.}
        \label{Planete_vs_DPI_2}
\end{figure*}

To test the implementation of the algorithm, we follow the method employed in Section 4.4 of \cite{Chambers02}. Here the authors compared their "wide binary integrator" with a reference integrator by computing the relative error on the semimajor axis calculation, $\Delta a/a$. Using the same approach, we perform a pure three-body simulation with a 10 \Mearth\ planet at 10 au of 1 \Msun\ primary star ($M_1$), which has a 0.5 \Msun\ companion ($M_2$) at 100 au binary separation. The integration lasts 10 Gyr with a variation in energy $\Delta E/E\sim10^{-10}$ and angular momentum $\Delta J/J\sim10^{-11}$, as we would expect given the symplectic nature of the integrator. Furthermore, the oscillation in planet semimajor axis and eccentricity do not show a secular trend, but they oscillate around a central value, meaning they are mostly round-off differences. We compare our results with a simulation run using the DPI library of \cite{TurriniCodice} with the same initial conditions and we computed $\Delta a/a = (a_\text{PAIRS}-a_\text{DPI})/a_{0}$ and $\Delta e = (e_\text{PAIRS}-e_\text{DPI})$ of the secondary star and planet (Fig. \ref{Planete_vs_DPI_1}). Given that both $\Delta a/a$ and $\Delta e$ are of order of magnitude $10^{-3}-10^{-4}$, comparable with the orders of magnitude obtained by \cite{Chambers02}, and also that the orbits of both secondary star and planet on the x-y plane overlap (Fig. \ref{Planete_vs_DPI_2}, data points from our code are displayed in orange while data points using the DPI library are displayed in blue), we consider our implementation accurate enough for the scope of our simulations. 

\section{Pure three-body simulations}\label{sec:set0}

\subsection{Initial conditions}

After testing the correct implementation of the algorithm (see Appendix \ref{sec:Nbody_testing}), we evaluate its accuracy and validity for different system architectures by running a grid of 5000 pure three-body simulations. The parameter space is randomly sampled as detailed in Table \ref{tab:initialconditions0}. The planet's mass $M_{\text{p}}$ is fixed to 10 \Mearth, while the primary star's mass $M_1$ is fixed at 1 \Msun. We vary the secondary star's mass $M_2$ between 0.1 and 1, corresponding to a binary mass ratio $q = M_2/M_1$ in the range 0.1-1. The binary separation, \abin, spans from 10 au to 1000 au, and the binary eccentricity, \ebin, varies between 0 and 0.9. The planet's initial location, $a_{\text{p,0}}$, is randomly chosen between 1 au and the location of the truncation radius $R_{\text{trunc}}$ minus 1 au (where the truncation radius is calculated as in PaperI). The simulations run until 10 Gyr, with the integrator timestep set to 2.5\% of the planet's orbital period. 

\begin{table}[h]
    \caption{Initial conditions for our pure three-body simulations.}
    \label{tab:initialconditions0}
    \centering
    \begin{tabular}{l|l}
        \hline
        \hline
        \multicolumn{2}{c}{Binary parameters}\\
        \hline
          $M_1$ [\Msun] & 1  \\
          $M_2$ [\Msun] & $\mathcal{U}$[0.1, 1] \\
          \ebin\  & $\mathcal{U}$[0, 0.9] \\
          \abin\  [au] & $\mathcal{U}$[10, 1000] \\
          \hline
        \multicolumn{2}{c}{Planet parameters}\\
        \hline
          $M_{\text{p}}$ [\Mearth] & 10 \\
          $a_{\text{p,0}}$ & $\mathcal{U}$[1 au, $R_{\text{trunc}}$ - 1 au] \\
         \hline
    \end{tabular}
\end{table}

\subsection{Results}\label{subsec:results_simulations0}

With our pure three-body simulations, we assess the accuracy of our integrator by running a grid of 5000 simulations with initial conditions outlined in Table \ref{tab:initialconditions0}.

\begin{figure}[h!]
        \centering
        \includegraphics[width=0.9\hsize]{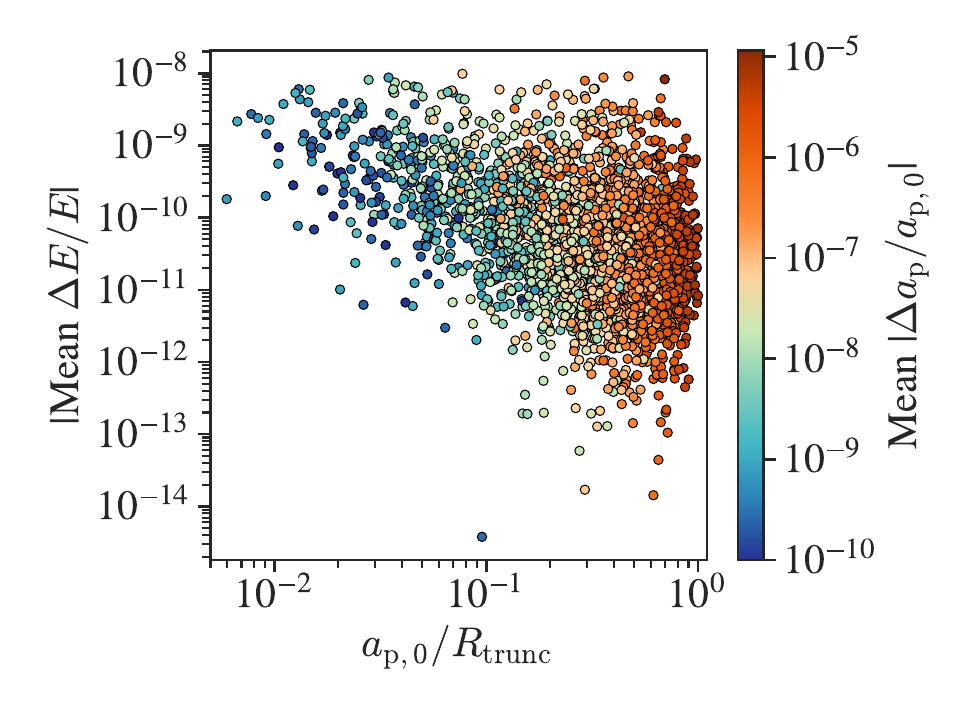}
        \vspace{-0.5cm}
        \caption{Pure three-body simulations: absolute value of the mean variation in energy, $\Delta E/E$, as a function of the planet’s initial location relative to the truncation radius, $a_{\text{p,0}}/R_\text{trunc}$. The color represents the mean amplitude of the variation in the planet’s semimajor axis, $\Delta a_\text{p}/a_\text{p,0}$. Only simulations where the planet was neither ejected nor accreted by the central star are shown, as such events would lead to significantly larger energy variations and semimajor axis errors.}
        \label{pure_Nbody_grid1}
\end{figure}

Figure \ref{pure_Nbody_grid1} shows the absolute value of the mean variation in energy $\Delta E/E$ as a function of the planet’s initial location in units of the truncation radius ($\mathcal{M}_1$). The color scale represents the amplitude of the oscillations in the planet’s semimajor axis, $|\Delta a_\text{p}/a_\text{p,0}|$. We only include simulations where the planet is neither ejected nor accreted by the central star, as these events would naturally lead to large variations in energy and semimajor axis, making it impossible to evaluate the integrator’s accuracy. For the simulations in which the planet remains bound, we find that the variation in energy stays, on average, below $10^{-8}$, while $|\Delta a_\text{p}/a_\text{p,0}|$ remains below $10^{-5}$. However, the latter increases as the planet approaches the disk truncation radius. Given the magnitudes of these oscillations, which are comparable, or better than, what obtained in the simulations of \cite{Chambers02}, we conclude that our integrator is sufficiently accurate to model the orbital evolution of different S-type architectures.

\begin{figure}[h!]
        \centering
        \includegraphics[width=0.9\hsize]{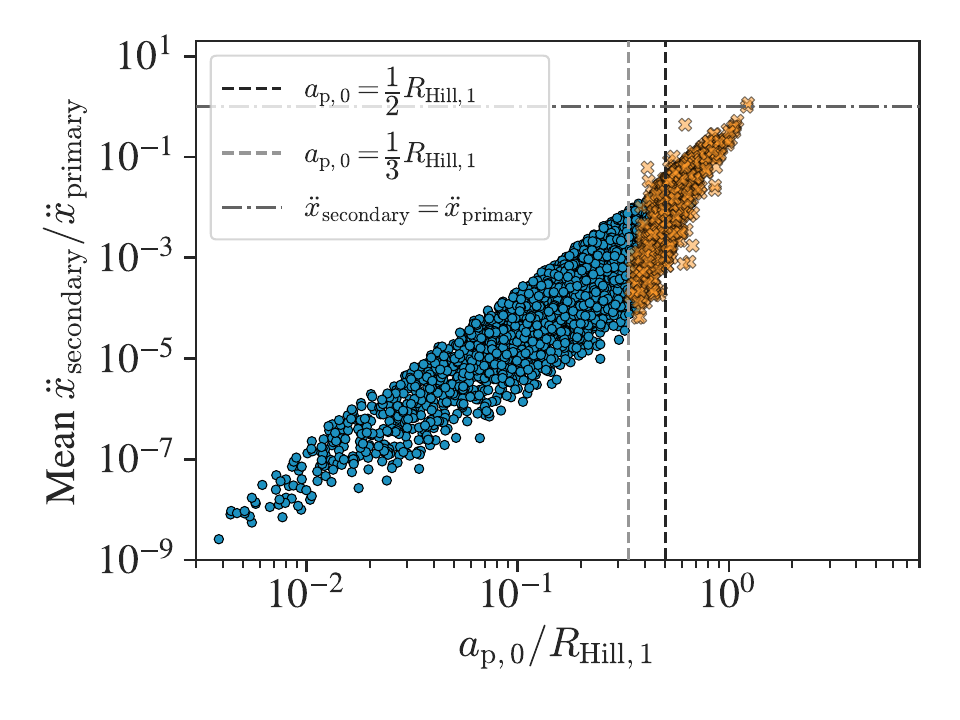}
        \vspace{-0.5cm}
        \caption{Pure three-body simulations: initial planet location relative to the primary star’s Hill radius, $a_{\text{p,0}}/R_\text{Hill,1}$, versus the mean value of the planet’s acceleration due to the secondary star ($\ddot{x}_{\text{secondary}}$) normalized by the acceleration due to the primary star ($\ddot{x}_{\text{primary}}$). The dashed gray and black  lines indicate orbital distances of $1/3$ and $1/2$ of $R_\text{Hill,1}$, respectively. The  dotted-dashed gray line indicates where $\ddot{x}_{\text{secondary}} = \ddot{x}_{\text{primary}}$. Blue circle markers represent planets that remain in the system, orange crosses indicate planets that are lost due to either ejection or instability due to a close encounter with the central star. We note that some blue markers are covered by the orange ones.}
        \label{pure_Nbody_grid3}
\end{figure}

\begin{figure}[h!]
        \centering
        \includegraphics[width=0.9\hsize]{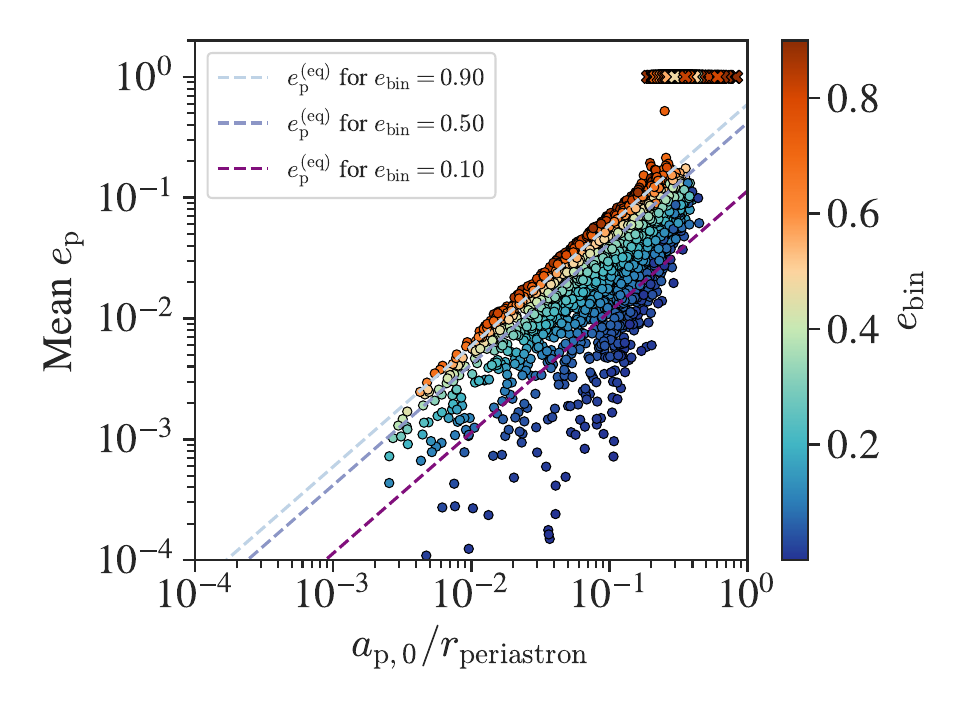}
        \includegraphics[width=0.9\hsize]{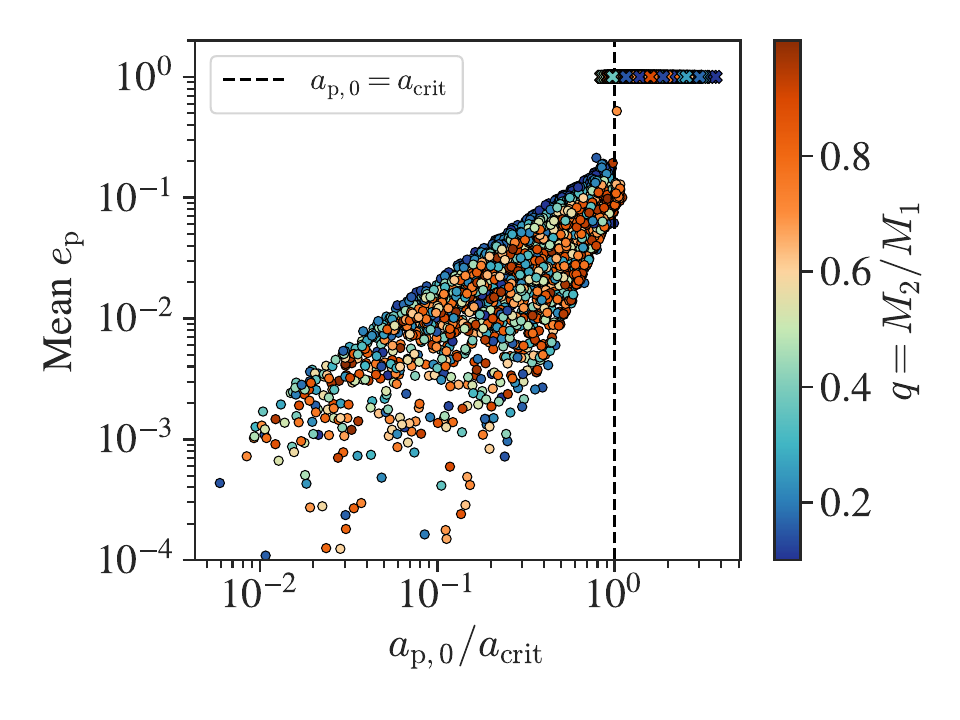}
        \vspace{-0.5cm}
        \caption{Pure three-body simulations: initial planet location relative to the binary periastron distance, $a_{\text{p,0}}/r_\text{periastron}$, versus mean planet eccentricity (top panel), and initial planet location relative to the critical semimajor axis, $a_{\text{p,0}}/a_\text{crit}$, versus mean planet eccentricity (bottom panel). In the top panel, the color represents the binary eccentricity and the purple dashed lines indicate the equilibrium eccentricity $e_\text{p}^\text{(eq)}$ \citep[Eq. (14) of][]{Mardling2007} for three binary eccentricities. In the bottom panel, the color represents the binary mass ratio and black dashed line marks the critical semimajor axis of \cite{Quarles2020}. Circle markers represent planets that remain in the system, crosses indicate planets that are lost due to either ejection or instability due to a close encounter with the central star.}
        \label{pure_Nbody_grid2}
\end{figure}

Including in our analysis also the systems where planets are lost due to ejection or instability due to a close encounter with the central star, we find that these planets are initially located at $a_\text{p,0} > R_{\text{Hill,1}}/3$ (i.e., $\mathcal{M}_2 > 1/3$) and, beyond $R_{\text{Hill,1}}/2$, the number of lost planets surpasses that of retained planets. Indeed, we find that 33.3\% of planets that are initially located between 1/3 and 1/2 of $R_{\text{Hill,1}}$ become unstable and this number increases to 84.4\% for planets that are initially  located beyond 1/2 of $R_{\text{Hill,1}}$. This trend is illustrated in Figure \ref{pure_Nbody_grid3}, where we show the mean value of the planet’s acceleration due to the secondary star ($\ddot{x}_{\text{secondary}}$) normalized by the acceleration due to the primary star ($\ddot{x}_{\text{primary}}$), as a function of $\mathcal{M}_2$. Blue circles indicate the planets that remain bound to their binary, while orange crosses indicate the planets that are lost from the system. We find that $\ddot{x}_{\text{secondary}}$ increases as the planet moves away from its host star. This is expected, as the planet is approaching the secondary star, it experiences a stronger gravitational influence with the companion, leading to a corresponding increase in its eccentricity and eventually ejection or instability due to a close encounter with the central star. The increase in planet eccentricity is shown in Figure \ref{pure_Nbody_grid2}. In its top panel we plot the mean planet eccentricity as a function of the planet location in units of the binary periastron distance $a_\text{p,0}/r_\text{periastron}$. The color bar shows the binary eccentricity, circle markers indicate the planets that remain bound to their binary, while crosses indicate the planets that are lost from the system. As for our set A simulations, we find that as the eccentricity of the binary increases, the mean planet eccentricity also increases. This dependence is described by Eq. (14) of \cite{Mardling2007}, which is shown, for three binary eccentricities, as purple lines in the top panel of Figure \ref{pure_Nbody_grid2}. 

In the bottom panel of Figure \ref{pure_Nbody_grid2} we plot the mean planet eccentricity against the planet’s initial location in units of the critical semimajor axis, $a_\text{crit}$, defined in Eq. \eqref{acrit}. Circles and crosses indicate bound and unbound planets, respectively. While there was a clear dependence of the planet eccentricity from the binary eccentricity, there is no strong dependence from the binary mass ratio, as shown by the color-coding of this bottom panel. Since the critical semimajor axis marks the outermost stable S-type orbit, planets initially placed beyond this threshold are ultimately lost from the system in the majority of cases. 
Quantitatively, from our simulations we find that 95.7\% of planets that are initially placed beyond $a_\text{acrit}$ become unbound. Conversely, only 3.9\% of planets that initially placed inside $a_\text{acrit}$ become unbound. This is also shown in Figure \ref{pure_Nbody_grid3}, where we plot the planet location as a function of the critical semimajor axis. Once again, circles and crosses indicate bound and unbound planets, respectively, and the dashed green line shows where $a_\text{p,0}=a_\text{acrit}$. 

\begin{figure}[h!]
        \centering
        \includegraphics[width=0.8\hsize]{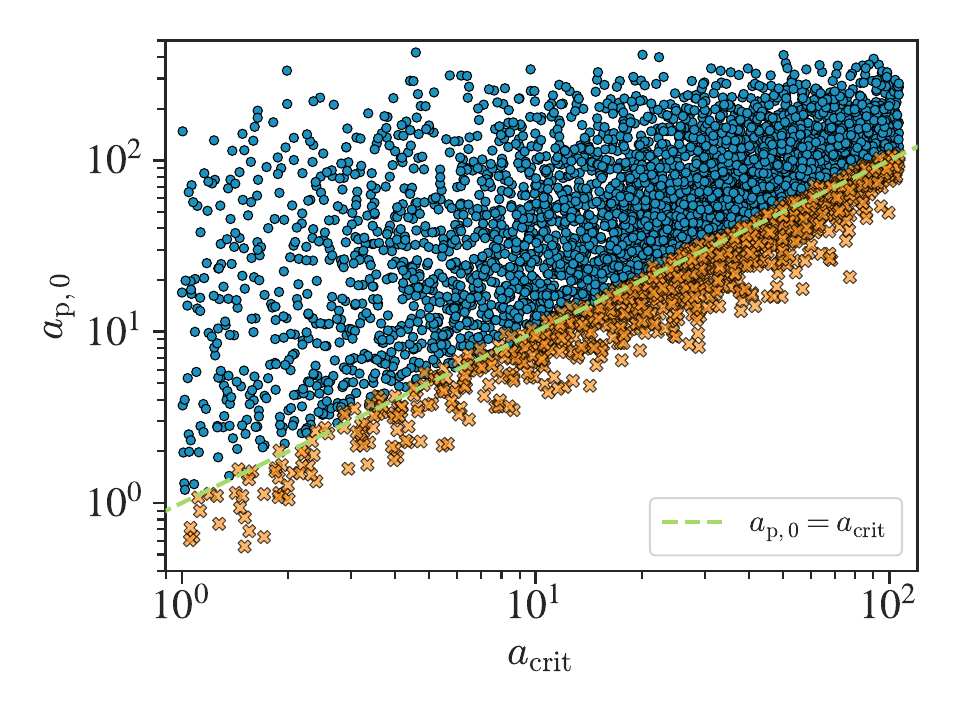}
        \vspace{-0.5cm}
        \caption{Pure three-body simulations: critical semimajor axis defined by Eq. \eqref{acrit} versus initial planet location. The  dashed green line indicates where $a_{\text{p,0}} = a_\text{crit}$. Blue circle markers represent planets that remain in the system, orange crosses indicate planets that are lost due to either ejection or instability due to a close encounter with the central star.}
        \label{pure_Nbody_grid3}
\end{figure}

\section{In-situ formation around the secondary star}\label{sec:circumsecondary}

Using the same setup as simulation set A (see left column of Table \ref{tab:initialconditionsABC}) and the same sampling of the parameter space, we performed a grid of in situ single-embryo planet formation simulations around the secondary, less massive, star, accounting for the scaling of the mass of the disk with the mass of the central star. 

The truncation radius of the circumsecondary disk, $R^{'}_\text{trunc}$, is computed as in Paper I and for the analysis we used the the critical stability boundary expression defined in Eq. \eqref{acrit} together with the metrics $\mathcal{M}^{'}_1$ and $\mathcal{M}^{'}_2$. They are equivalent of the metrics $\mathcal{M}_1$ and $\mathcal{M}_2$, but in the context of the circumsecondary disk, and they are defined as:
\begin{equation}\label{metric1_}
        \mathcal{M}^{'}_1 = \dfrac{a_\text{p,0}}{R^{'}_\text{trunc}}\quad ; \quad \mathcal{M}^{'}_2 = \dfrac{a_\text{p,0}}{R_\text{Hill,2}}
\end{equation}

where $R_\text{Hill,2}$ is the Hill radius of the secondary star and is defined as:
\begin{equation}\label{Hill_sphere_2}
    R_\text{Hill,2} = a_\text{bin}\cdot(1-e_\text{bin})\cdot\left[\dfrac{M_2}{3\cdot(M_1 + M_2)}\right]^{1/3}.
\end{equation}

From our simulations, 88.1\% systems retain their planets, while 4.9\% systems lose their planets due to ejection, and 2.8\% systems become unstable due to a close encounter with the central star. We also have 4.2\% of systems that, given their truncation radius and mass of gas before truncation, had a mass of gas after truncation below $10^{-5}$\Msun\ and therefore the embryos could not be initialized. In the following discussion we are going to focus on the systems that we were able to initialize given that there was a sufficient amount of gas after truncation. 

\begin{figure}[h!]
        \centering
        \includegraphics[width=0.9\hsize]{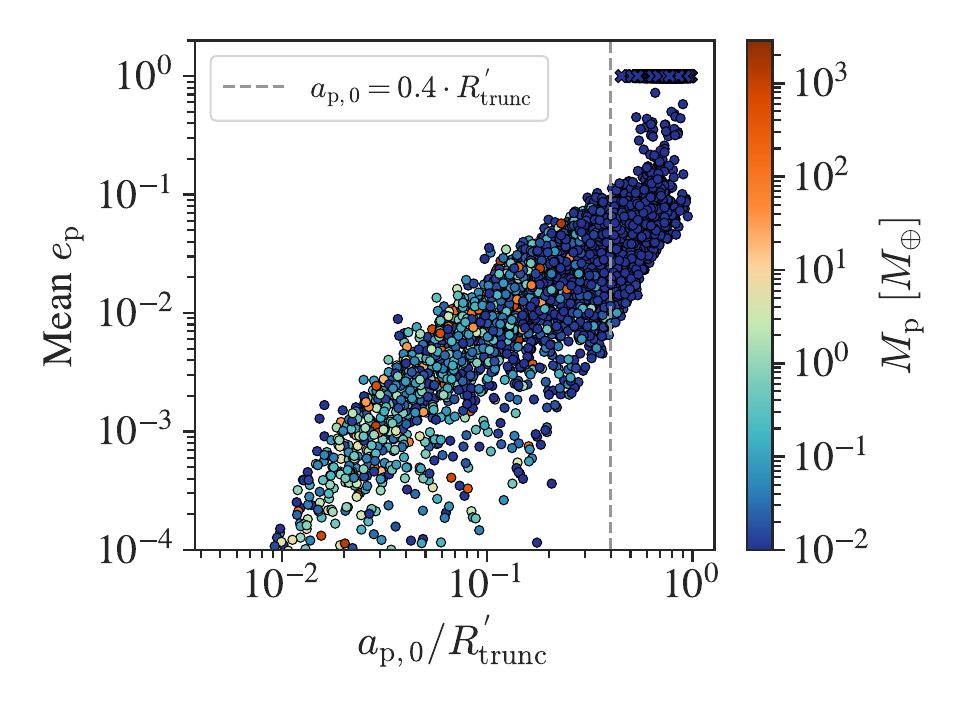}
        \includegraphics[width=0.9\hsize]{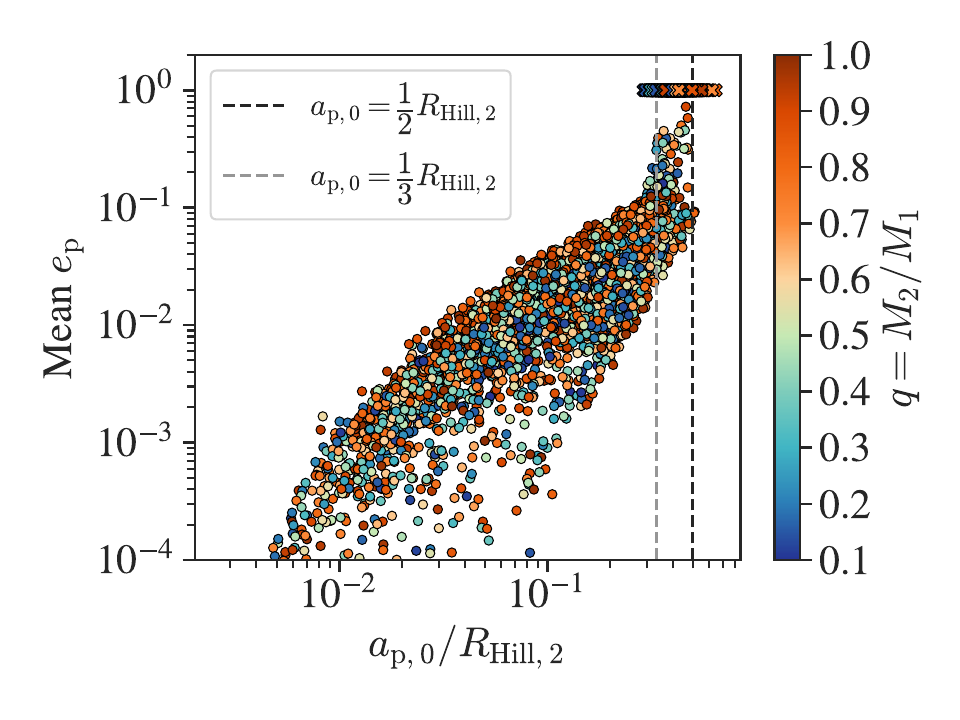}
        \vspace{-0.5cm}
        \caption{Circumsecondary in situ formation: Initial planet location relative to the disk truncation radius ($a_{\text{p,0}}/R^{'}_\text{trunc}$, top), the primary star’s Hill radius ($a_{\text{p,0}}/R_\text{Hill,2}$, bottom). Top panel: Color bar shows the final planet mass and the gray dashed line marks $0.4\cdot R^{'}_\text{trunc}$. Bottom panel:  Color bar shows the binary mass ratio and the gray and black dashed lines mark $1/3$ and $1/2\cdot R_\text{Hill,2}$. Circles denote surviving planets, while crosses indicate planets lost by ejection or instability from close encounters with the central star.}
        \label{circumsecondary_insitu_grid1}
\end{figure}

Figure \ref{circumsecondary_insitu_grid1} shows the mean planet eccentricity as a function of $\mathcal{M}^{'}_1$, color-coded by the final planet mass (top panel), and as a function of $\mathcal{M}^{'}_2$, color-coded by the binary mass ratio (bottom panel). Circle markers represent systems that retain their planets, while cross markers indicate planets lost due to ejection or instabilities due to a close encounter with the central star, respectively. Just as in the case of the circumprimary disk, we find that planets initially located beyond 1/3 of $R_\text{Hill,2}$ start to be lost and the frequency of losses increases as planets approach 1/2 of $R_\text{Hill,2}$. Namely, 51.7\% of systems with  $1/3<\mathcal{M}^{'}_2<1/2$ and 94.9\% of systems with $\mathcal{M}^{'}_2>1/2$ experience the loss of their planet. We also compare our results with the dynamical stability boundary defined by \citet{Quarles2020} and we find that only 3.7\% of planets located within $a_{\text{crit}}$ are lost, compared to 71.5\% of those beyond it. 

\begin{figure}[h!]
        \centering
        \includegraphics[width=0.87\hsize]{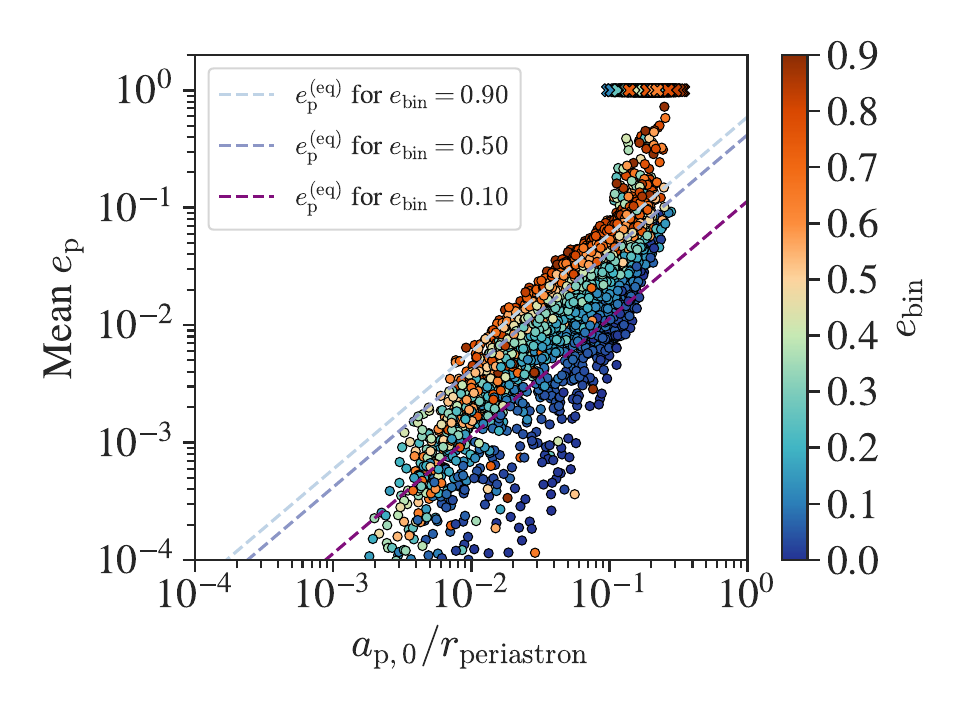}
        \vspace{-0.5cm}
        \caption{Circumsecondary in situ formation: Initial planet location relative to the binary periastron distance ($a_{\text{p,0}}/r_\text{periastron}$) versus mean planet eccentricity. The color bar shows the binary eccentricity and purple dashed lines indicate the equilibrium eccentricity $e_\text{p}^\text{(eq)}$ \citep[Eq. (14) of][]{Mardling2007} for three binary eccentricities. Circles denote surviving planets, while crosses indicate planets lost by ejection or instability from close encounters with the central star.}
        \label{circumsecondary_insitu_grid2}
\end{figure}

From the color-coding of the bottom panel of Figure \ref{circumsecondary_insitu_grid1} we don't find any correlation between the planet eccentricity and the binary mass ratio, however there is a dependence from the binary eccentricity. This is shown in Figure \ref{circumsecondary_insitu_grid2}, where we plot the mean planet eccentricity as a function of the planet location relative to the binary periastron distance, color-coded by binary eccentricity. Stable and lost planets are shown with circles and crosses, respectively. This correlation results from the relation between the planet equilibrium eccentricity and the binary eccentricity, as defined by Eq. (14) of \cite{Mardling2007}. We plot this relation for three values of the binary eccentricity as dashed purple lines in the same Figure \ref{circumsecondary_insitu_grid2}.

\begin{figure}[h!]
        \centering
        \includegraphics[width=0.9\hsize]{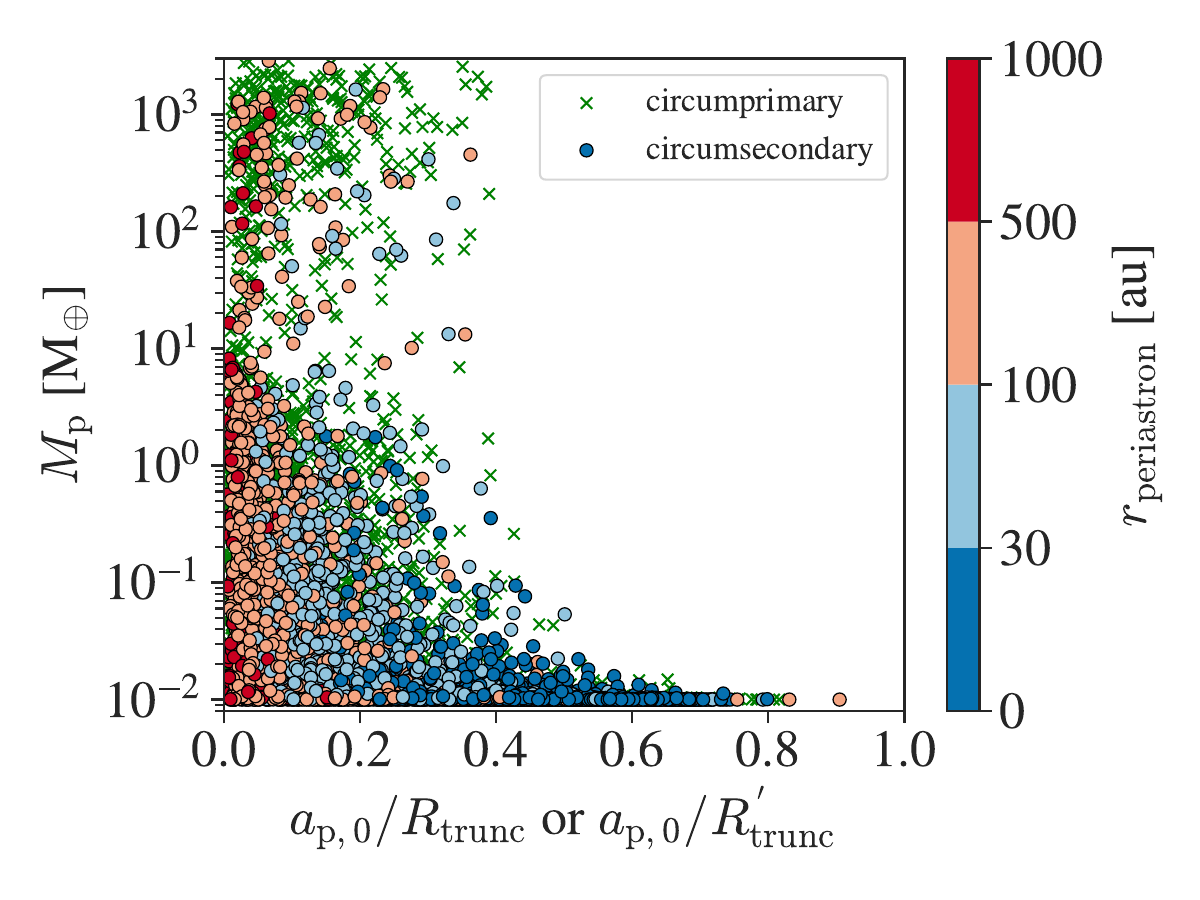}
        \vspace{-0.5cm}
        \caption{Circumsecondary in situ formation: Initial planetary position relative to the disk truncation radius, $a_{\text{p,0}}/R_\text{trunc}$ or $a_{\text{p,0}}/R^{'}_\text{trunc}$, versus final planet mass. The color bar shows the binary periastron distance. Circle markers represent stable planets from circumsecondary simulations presented in this section, while green crosses represent stable planets from set A simulations.}
        \label{circumsecondary_insitu_grid3}
\end{figure}

In Figure \ref{circumsecondary_insitu_grid3} we plot the final planet mass as a function planet location relative to the disk truncation radius, color-coded with binary periastron distance. Circle markers show the circumsecondary simulations, while green crosses show the set A simulations presented in Section \ref{subsec:results_simulationsA}, namely, the circumprimary simulations. We find that planets that survive in the region $\mathcal{M}^{'}_1 > 0.4$ are inefficient at accreting material. As a matter of fact, no planets with mass above the mass of Mars are formed beyond this threshold (as also shown by the color bar in the top panel of Figure \ref{circumsecondary_insitu_grid1}). 

\begin{figure}[h!]
        \centering
        \includegraphics[width=0.9\hsize]{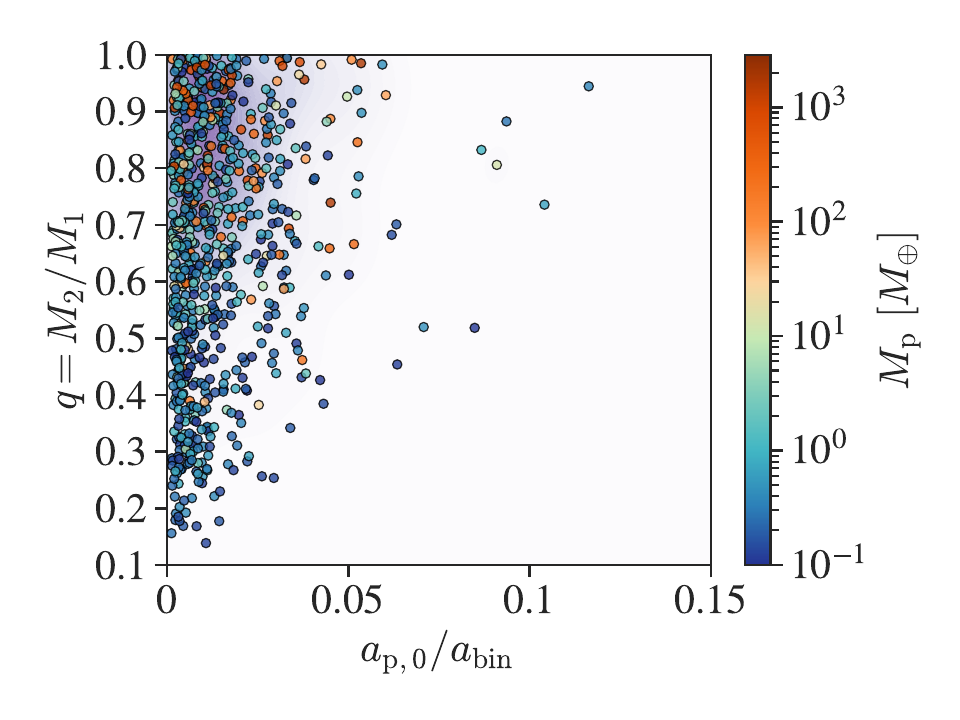}
        \includegraphics[width=0.9\hsize]{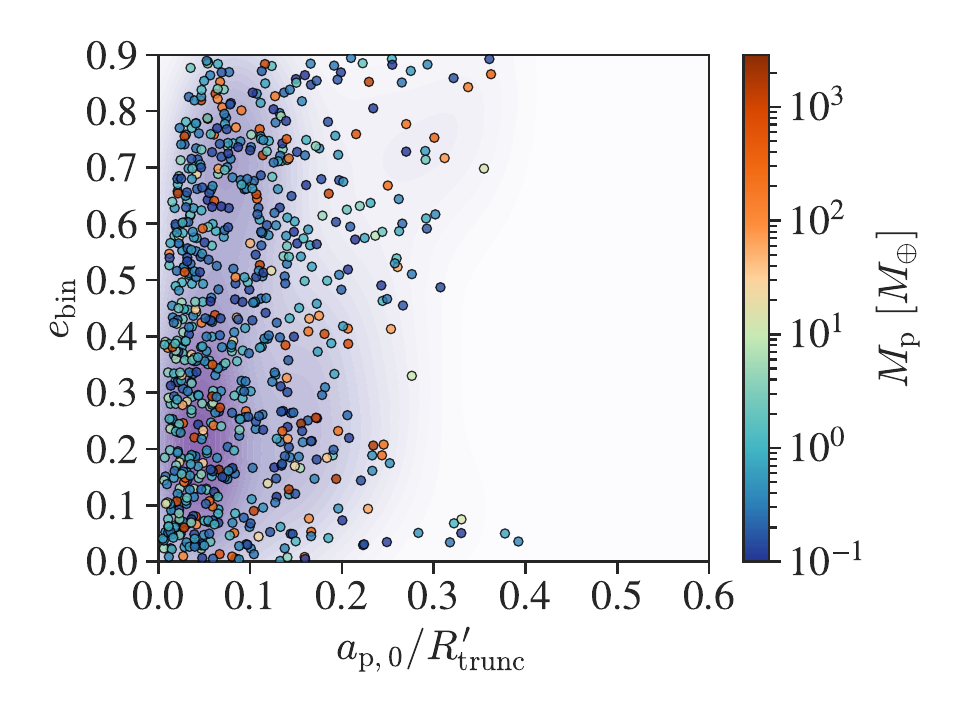}
        \vspace{-0.5cm}
        \caption{Circumsecondary in situ formation: Initial planetary position relative to the binary separation, $a_{\text{p,0}}/a_\text{bin}$, versus the binary mass ratio (top panel) and initial planetary position relative to the disk truncation radius, $a_{\text{p,0}}/R^{'}_\text{trunc}$, versus the binary eccentricity (bottom panel). We show planets that became at least more massive than Mars and their color reflects their final planet mass. The shaded contours highlight regions of highest planet density, obtained through a two-dimensional kernel density estimation (KDE) considering planets with masses above 10 \Mearth.}
        \label{circumsecondary_insitu_grid4}
\end{figure}

On the other hand, planets that form inside 0.4 $R^{'}_\text{trunc}$, and especially inside 0.3 $R^{'}_\text{trunc}$, can grow significantly, reaching masses of up to $\sim 6$ \Mj. However, compared to the formation around the primary star, we form less massive planets and especially less giant planets. Indeed, the mean mass of the planets located within 30\% of the disk around the primary star was $\sim$112 \Mearth, while the mean mass of the planets located within 30\% of the disk around the secondary star is $\sim 24$ \Mearth. As an example, 6.3\% of systems from set A formed a planet with $M_\mathrm{p}>100$ \Mearth, while only 1.5\% of systems in the circumsecondary case formed such planets. If we look at what type of binary architectures allowed the formation of these planets, we find that the majority has $r_\text{periastron}$ > 100 au. Furthermore, in Figure \ref{circumsecondary_insitu_grid4} we plot the binary mass ratio as a function of planet location relative to the binary separation, $a_{\text{p,0}}/a_\text{bin}$ (top panel) and binary eccentricity as a function of $\mathcal{M}^{'}_1$ (bottom panel). In both panels we plot only the planets that grew at least as massive as Mars, with the color bar shows their final mass, and the purple contour region highlights the highest density of planets above 10 \Mearth. We find that rather than a dependence on the binary eccentricity alone, the final mass that a planet can reach depends on how the binary eccentricity, as well as the other binary parameters (Paper I) enter into the periastron distance and disk truncation prescription. However, compared to the circumprimary case, where the final planet mass was not depending from the binary mass ratio (top panel of Figure \ref{insitu_grid4}), now that we are modeling the formation around the less massive star, the majority of planets with mass above 10 \Mearth\ are formed when $q$ > 0.6.

\end{appendix}

\end{document}